\documentclass[prd,twocolumn,nofootinbib,showpacs]{revtex4}
\usepackage[final]{graphicx}
\usepackage{bm}
\usepackage{amssymb}
\usepackage{natbib}
\usepackage{amsmath}

\newcommand{\orbit}{z^{\alpha}}
\newcommand{\geo}{z_G^{\alpha}}
\newcommand{\diff}[2]{\frac{d #1}{d #2}}
\newcommand{\pdiff}[2]{\frac{\partial #1}{\partial #2}}
\newcommand{\ddiff}[2]{\frac{d^2 #1}{\ d #2^2}}
\newcommand{\pddiff}[2]{\frac{\partial^2 #1}{\ \partial #2^2}}
\newcommand{\Chr}[3]{\Gamma^{#1}_{\ #2#3}}
\newcommand{\av}[1]{\langle #1\rangle}

\begin{document}

\title{Osculating orbits in Schwarzschild spacetime, with an
  application to extreme mass-ratio inspirals} 
\author{Adam Pound and Eric Poisson} 
\affiliation{Department of Physics, University of Guelph, Guelph,
  Ontario, N1G 2W1} 
\date{December 18, 2007}

\begin{abstract}
We present a method to integrate the equations of motion that govern
bound, accelerated orbits in Schwarzschild spacetime. At each instant
the true worldline is assumed to lie tangent to a reference geodesic,
called an osculating orbit, such that the worldline evolves smoothly
from one such geodesic to the next. Because a geodesic is uniquely
identified by a set of constant orbital elements, the transition
between osculating orbits corresponds to an evolution of the
elements. In this paper we derive the evolution equations for a
convenient set of orbital elements, assuming that the force acts only
within the orbital plane; this is the only restriction that we impose
on the formalism, and we do not assume that the force must be
small. As an application of our method, we analyze the relative motion
of two massive bodies, assuming that one body is much smaller than the
other. Using the hybrid Schwarzschild/post-Newtonian equations of
motion formulated by Kidder, Will, and Wiseman, we treat the
unperturbed motion as geodesic in a Schwarzschild spacetime with a
mass parameter equal to the system's total mass. The force then
consists of terms that depend on the system's reduced mass. We
highlight the importance of conservative terms in this force, which
cause significant long-term changes in the time-dependence and phase
of the relative orbit. From our results we infer some general
limitations of the radiative approximation to the gravitational
self-force, which uses only the dissipative terms in the force.  
\end{abstract}
\maketitle


\section{Introduction}

\subsection{Orbital motion in curved spacetime}

Analysis of accelerated orbits in curved spacetime has historically
focused on the post-Newtonian regime (see
Ref.~\cite{300_years,Blanchet,PN_review2} for general reviews of the 
post-Newtonian formalism), since observations of orbital motion have
historically been limited to weak-field systems such as the
solar neighborhood and binary pulsars. However, the advent of
gravitational-wave astronomy has recently necessitated an analysis of
accelerated orbits in strongly-curved spacetimes. The primary examples
of such orbits are extreme mass-ratio inspirals (EMRIs), in which a
small compact body of mass $m$ spirals into a supermassive black hole
of mass $M\gg m$. Such systems promise to be excellent sources of
gravitational waves for the space-based detector LISA
\cite{LISA}. However, accurate predictions of the emitted waveforms
must account for the effect of the compact body's gravitational field
on its own motion. The compact body induces a metric perturbation 
$h_{\alpha\beta}= (m/M) h^{(1)}_{\alpha\beta} 
+ \mathcal{O}(m/M)^2$. Although the motion of the particle may be
described as a geodesic in the perturbed spacetime, it is more simply
treated as an accelerated worldline in the background spacetime of the
unperturbed black hole. The cause of the acceleration is thus
interpreted as a gravitational self-force derived from a regularized
form of the field $h_{\alpha\beta}$. This force was first formally
calculated to first order in $m/M$ by Mino, Sasaki, and Tanaka
\cite{Mino_etal}, and later by Quinn and Wald \cite{QuinnWald} 
(see Ref.~\cite{self_force_review} for a review of recent
developments). Other possible effects on the inspiraling particle,
such as tidal perturbations of the central black hole, spin-orbit and
spin-spin couplings, electromagnetic interactions, and so on, can also 
be treated as forces acting on the body.  

Although significant progress has been made in calculating these 
effects (see Ref.~\cite{Drasco_review} for a recent review of work on
EMRIs), there has been no attempt to formulate a general method of 
determining and characterizing the resulting motion. Implementing the
first-order gravitational self-force brings a particular difficulty:
The self-force on a particle is a functional of the particle's
worldline, which for the first-order calculation is assumed to be a
geodesic. However, the true motion is never geodesic, because of the
self-force. Thus, the effect of the self-force must somehow be
determined with reference to a fictitious geodesic worldline.  

In this paper we present a method to integrate the equations of motion
that govern accelerated motion in Schwarzschild spacetime. The method
can be used for a wide class of perturbing forces; the only
restrictions are that the force must keep the orbital motion bounded
between a minimum and a maximum radius (the method is not suitable for
the final portion of an orbit that plunges into the black hole), and
that it must be acting within the plane of the orbit (although the
method could be easily extended to accommodate non-planar
motion). Within these restrictions the force is 
arbitrary, and in particular, it is not assumed to be small. Our
method is a relativistic extension of the traditional method of
osculating orbits, also called the method of variation of constants,
in Newtonian celestial mechanics (see, e.g., Refs.~\cite{Taff,
Beutler}). In this method the true worldline $z(\lambda)$ is 
taken to lie tangent to a geodesic $z_G(\lambda)$ at each value of the 
orbital parameter $\lambda$, such that the true orbit moves smoothly
from one geodesic to the next. The instantaneously tangential
geodesics are referred to as osculating orbits (meaning ``kissing
orbits''). A geodesic is characterized by a set of constants $I^A$,
called \textit{orbital elements}, and the transition between
osculating orbits corresponds to changes in these elements; thus, the
method of osculating orbits amounts to parametrizing the true
worldline as an evolving geodesic with dynamical orbital elements
$I^A(\lambda)$.

Because it explicitly determines the position and velocity of a 
tangential geodesic at each instant, this method explicitly provides
the information necessary to calculate the first-order gravitational
self-force at each instant. Our method is therefore very well suited
to the gravitational self-force problem. It also has several more
general advantages. First, because the orbital elements 
are constant on a geodesic, the method clearly separates perturbative 
from non-perturbative effects. (Throughout this paper the accelerated
motion of the particle is referred to as a perturbation of the
geodesic motion. However, this is only to distinguish effects of
acceleration from effects on a geodesic; the ``perturbation" need not
be small.) Second, although the orbital elements are equivalent to the
set of initial conditions, they are typically chosen so as to provide
direct geometric information about the orbit. If the perturbing force
is very weak, then the perturbed orbit will lie very close to a
geodesic for a long period of time, and changes in the orbital
elements will characterize changes in the geometry of the orbit. Thus,
although our method is exact, it is perhaps most useful in the context
of small perturbations. Third, the orbital elements divide into two
classes. The first class includes the \textit{principal} orbital
elements; these are equivalent to constants of the motion such as
energy and angular momentum, and they determine the geodesic on which
the particle is moving. The second class includes the
\textit{positional} orbital elements, which determine the particle's
initial position on the selected geodesic, as well as the geodesic's
spatial orientation. Generally speaking, long-term changes in the
principal orbital elements are produced by dissipative terms in the
perturbing force, while long-term changes in the positional elements
are produced by conservative terms. Thus, this division into two
classes allows one to easily separate conservative from dissipative
effects of the perturbing force.  

%
We note that this general idea of characterizing orbital evolutions
by changes in the ``constants'' of motion has been used frequently in  
analyzing the effects of radiation reaction. Such analyses have
typically focused on changes in the principal elements alone,
neglecting the changes in positional elements, and rarely mentioning
the general framework of osculating orbits. However, there have been
at least two notable generalizations of the method of osculating
orbits from Newtonian to relativistic mechanics: the adaptation of the
method by Damour et~al.\ to post-Newtonian binary systems
\cite{Damour, Damour2004}, and the formulation proposed by Mino for
orbits around a Kerr black hole \cite{Mino2005}. The formulation by
Damour et~al.\ is complete and easy to implement, but it is limited to
the post-Newtonian regime. Mino's formulation is valid for arbitrary
bound orbits in Kerr, and it was undoubtedly useful for Mino's own
purposes, but we believe that a concrete implementation of his method
would not be very practical. The reason is that Mino expresses the
orbits as formal Fourier expansions with unknown convergence behavior,
in terms of coefficients that would be difficult to calculate in
practice. It may well be that the complexity of geodesics in Kerr make
a more practical parametrization impossible, but as we shall
demonstrate in this paper, we can do much better for orbits in
Schwarzschild spacetime. Given the limitations of previous work, we
believe that it is timely to present here a practical formulation of
the method of osculating orbits for bound motion in Schwarzschild
spacetime. We shall first present an outline of the general method in
relativistic mechanics and its connection to the traditional method in
Newtonian mechanics, and we shall next specialize the method to the
case of bound orbital motion in Schwarzschild spacetime.   
%
%

\subsection{Test case: post-Newtonian binaries}
\label{PN binaries}

We demonstrate the usefulness of our method by applying it to the 
relatively simple system of two compact bodies of mass $m_1$ and
$m_2\gg m_1$ in the post-Newtonian regime. The equations of motion for
the spatial positions $x^a_1$ and $x^a_2$ of the bodies have been
determined in harmonic coordinates to 3.5PN order (i.e. of order
$(v/c)^7$ beyond the Newtonian description)
\cite{Blanchet}. Conservative terms appear at 1PN, 2PN, and 3PN
orders, and dissipative terms appear at 2.5PN and 3.5PN orders. Since
the essential features of the problem are already present at 2.5PN
order, we truncate the equations at that order for simplicity. These
equations are valid for arbitrary mass ratios, but we focus on the
extreme case in order to link our results to the self-force problem.  

In order to analyze this system of equations with our method of
osculating orbits, we use the hybrid equations of motion constructed
by Kidder, Will, and Wiseman~\cite{Kidder}. These equations take the
schematic form  
\begin{equation}
\ddiff{x^a}{t}=-\frac{M}{r^2}\left( 1 + \mbox{\sc{Schw}}
+ \mu \mbox{\sc{pf}} \right). 
\end{equation}
The spatial separation vector $x^a = x^a_1 - x^a_2$ connects the two
bodies, and $M = m_1 + m_2 $ and $\mu = m_1 m_2/M$ are respectively
the total mass and reduced mass of the system. The terms in
$\mbox{\sc{Schw}}$ are the exact relativistic corrections to Newton's
law in a Schwarzschild spacetime of mass $M$, so that
$\ddiff{x^a}{t}=-\frac{M}{r^2}\left(1+\mbox{\sc{Schw}}\right)$ is the
exact equation for a test particle in that spacetime. The terms in
$\mu \mbox{\sc{pf}}$ are those in the post-Newtonian expansion that depend  
explicitly on the reduced mass of the system ($\mbox{\sc{pf}}$ stands for
``perturbing force''). Since the extra terms introduced within
$\mbox{\sc{Schw}}$ are of 3PN order and higher, the hybrid equations
remain correct at 2.5PN order. However, they differ from the usual 
post-Newtonian equations in that they become exact in the test-mass
limit $\mu \to 0$. This allows us to apply our method to the
post-Newtonian system by taking our osculating orbits to be geodesics
in the fictitious Schwarzschild spacetime of mass $M$, and by deriving
our perturbing force from $\mu \mbox{\sc{pf}}$.  

The force derived in this way is a form of the gravitational
self-force, since it is produced by finite-mass effects. However, it 
differs nontrivially from the post-Newtonian limit of the relativistic 
self-force: First, the self-force is a gauge-dependent quantity which
is typically calculated in the Lorenz gauge, while the hybrid
equations of motion are derived within the harmonic gauge of
post-Newtonian theory. Second, the Lorenz gauge ensures that the
coordinates of the small body are defined in relation to the system's
center of mass \cite{Eric_Steve}, while here we use coordinates
relative to the large mass. And third, our geodesics are in a
fictitious Schwarzschild spacetime of mass $M = m_1 + m_2$ and not in
the background spacetime of the second body (of mass $m_2$). The last
two differences could be easily removed by formulating an alternative
set of hybrid equations, but the gauge difference cannot be easily
dealt with.    

Given these differences, our method of osculating elements is used in 
this paper primarily as a practical means to integrate the hybrid
equations of motion. Nevertheless, the perturbing force that we derive
and the gravitational self-force share many essential features. In
particular, the self-force can be expected to have conservative terms
at 0PN (the Newtonian level), 1PN, and 2PN orders, etc., and
dissipative terms at 2.5PN (corresponding to quadrupole radiation) and
3.5PN orders, etc.; our perturbing force has exactly the same
features, except for the Newtonian correction, which is implicitly
accounted for by working in terms of total and reduced masses. Thus,
we can hope to draw some reasonable conclusions about the action of
the gravitational self-force from our simplified analysis.   

%
Our focus will be on detailing the limitations and ambiguities of 
two approximation schemes, following our analysis of the
post-Newtonian electromagnetic self-force in Refs.~\cite{our_paper,
other_paper}. The first scheme of interest lies within the broad class 
of adiabatic approximations, which rest on the assumption that the
accelerated orbit deviates only ``slowly'' from the geodesic orbit. In
particular, they commonly assume that any period of the motion is much
shorter than the radiation-reaction timescale of the inspiral,
allowing one to eliminate irrelevant short-term oscillations and keep
only secular effects. Based on this assumption, an explicit
implementation of such an approximation will typically involve some
type of averaging, either in the form of direct averaging of the
equations of motion or via a two-timescale expansion. For clarity, we
will refer to this averaging method, which is just a specific type of
adiabatic approximation, as a \emph{secular approximation}. Using the
hybrid equations of motion, we show in Sec.~III B that the secular
approximation introduces ambiguities in the choice of (a) initial
conditions and (b) the variable to be averaged over. Our results
suggest that different choices can significantly affect long-term
behavior, and our conclusion is that while the idea of a secular
approximation is attractive, the precise construction of one presents
significant difficulties. 
%

We shall also examine the (pseudo-adiabatic) \emph{radiative
approximation}, which uses the radiative (half-retarded minus
half-advanced) solution to the linearized Einstein equation. As shown
by Mino \cite{Mino}, the self-force calculated from the radiative
field approximately reproduces the long-term dissipative effects of
the true self-force. Largely based on this result, it was believed
that the radiative approximation would produce a valid adiabatic
approximation to the true evolution. This notion has led to a
confusing nomenclature in the literature, in which adiabatic and
radiative approximations are treated synonymously. Since the radiative
approximation introduces errors beyond those of an adiabatic
approximation \cite{our_paper, other_paper}, we find it misleading to
identify the two. We insist here that the radiative approximation is
logically distinct from the class of adiabatic approximations
introduced in the preceding paragraph. 

Due to its simplicity, the radiative approximation has been utilized 
by several groups in analyzing EMRIs
\cite{Nakano,Hughes}. Unfortunately, the radiative self-force neglects  
all conservative effects of the true self-force. In the framework of
osculating orbits, this translates into neglecting long-term
changes in the positional orbital elements. (Although Mino has given 
prescriptions for finding these long-term changes using only the
radiative self-force \cite{Mino, Mino2005}, his prescriptions are
highly ambiguous in practice \cite{my_thesis}.) As pointed out in
Ref.~\cite{Hughes}, the radiative approximation may have some utility
despite this error, and in particular, it may be sufficient to
generate templates for the detection of a gravitational-wave
signal. But it is unlikely that it will be sufficiently accurate for
reliable parameter estimation. Because of the potential usefulness of
the approximation, determining its limitations is quite important. In
this paper we find that neglecting conservative effects leads to
long-term errors in the phase and time-dependence of the orbit; this
agrees with and extends our earlier results \cite{our_paper,
other_paper}. The errors in the time dependence are of particular
importance, as they apply even to the evolution of the principal
orbital elements.   

\subsection{Organization of this paper}

In Sec.~\ref{general case} we introduce the general method of
osculating orbits. We then restrict our analysis in
Secs.~\ref{geodesics} and \ref{evolution} to bound planar orbits in
Schwarzschild spacetime. Section~\ref{geodesics} presents a
parametrization of bound geodesics in terms of five orbital elements,
and Sec.~\ref{evolution} uses the osculation condition to find
evolution equations for these orbital elements. In the second part of
our paper, we apply our method to the hybrid
Schwarzschild/post-Newtonian equations of motion, which are
presented in Sec.~\ref{hybrid}. The results of using a secular or
radiative approximation are then displayed and discussed in
Sec.~\ref{results}.  


\section{The method of osculating orbits}
\label{method}

\subsection{The osculation condition}
\label{general case}

We first consider the completely general situation of a point particle 
moving on an arbitrary worldline $\orbit(\lambda)$ parametrized by
$\lambda$. We define the acceleration $f^\alpha$, or force per unit
mass, acting on the particle via the equation of motion  
\begin{equation}\label{affine eq mot}
\ddot z^{\alpha}+\Chr{\alpha}{\beta}{\gamma} 
\dot z^{\beta} \dot z^{\gamma} = f^{\alpha},
\end{equation}
where an overdot indicates a derivative with respect to the proper
time $\tau$ on the worldline. The normalization condition 
$\dot z^{\alpha}\dot z_{\alpha} = -1$ implies the orthogonality
condition $f^{\alpha}\dot z_{\alpha} = 0$, which will be essential for
later calculations. The relation between $f^{\alpha}$ and the Newtonian
perturbing force is discussed in Appendix~\ref{Newtonian}.   

Using the relations $\dot z^{\alpha} =
\diff{\orbit}{\lambda}\dot\lambda$ and $\ddot z^{\alpha} =
\ddiff{\orbit}{\lambda}\dot\lambda^2+\diff{\orbit}{\lambda}\ddot\lambda$, 
the equation of motion becomes 
\begin{equation}
\ddiff{\orbit}{\lambda}+\Chr{\alpha}{\beta}{\gamma} 
\diff{z^{\beta}}{\lambda}\diff{z^{\gamma}}{\lambda} 
= f^{\alpha}\left(\diff{\tau}{\lambda}\right)^2
+ \kappa(\lambda)\diff{\orbit}{\lambda},
\label{eq mot}
\end{equation}
where $\kappa=-\ddot\lambda/\dot\lambda^2$. The first term on the 
right-hand side is due to the force acting on the particle, while the
second term is present whenever $\lambda$ is a non-affine parameter. 

Our goal is to transform the equation of motion~\eqref{eq mot} into 
evolution equations for a set of orbital elements $I^A$. That is, we
seek a transformation $\{\orbit,\dot z^{\alpha}\} \to I^A$. Letting 
$\geo(I^A,\lambda)$ be a geodesic with orbital elements $I^A$, the
\textit{osculation condition} states the following:  
\begin{eqnarray}
\orbit(\lambda) & = & \geo(I^A(\lambda),\lambda), 
\label{osc 1} \\
\diff{\orbit}{\lambda}(\lambda) & = & 
\pdiff{\geo}{\lambda}(I^A(\lambda),\lambda), 
\label{osc 2}
\end{eqnarray}
where the partial derivative in the second equation holds $I^A$ 
fixed. These two equations assert that at each value of $\lambda$ we
can find a set of orbital elements $I^A(\lambda)$ such that the
geodesic with those elements has the same position and velocity as the 
accelerated orbit. We can freely make this assertion because the
number of orbital elements is equal to the number of degrees of
freedom on the orbit. 

As a consequence of the osculation condition, all relations that are
obtained using only algebraic manipulations of coordinates and
velocities on a geodesic are also valid on the true orbit. However, it  
is important to note that $\kappa$ is altered by the acceleration of 
the worldline, because it involves second derivatives. Hence, an
expression for $\kappa(\lambda)$ that is valid on an osculating
geodesic will not be valid on the tangential accelerated
orbit. Nevertheless, $\ddot\lambda=0$ for an affine parameter
$\lambda$ on both orbits, so affine parameters remain affine.   

Now, combining the osculation condition with the equations of motion 
generates evolution equations for $I^A$. From Eq.~\eqref{osc 1} we
have that $\diff{\orbit}{\lambda}=\diff{\geo}{\lambda}$, which implies 
$\diff{\orbit}{\lambda} = \pdiff{\geo}{\lambda} 
+ \pdiff{\geo}{I^A}\diff{I^A}{\lambda}$, where the index $A$ is summed
over. Comparing this result with Eq.~\eqref{osc 2}, we find 
\begin{equation}
\pdiff{\geo}{I^A}\diff{I^A}{\lambda} = 0.
\label{diffI 1}
\end{equation}
Furthermore, $\geo$ satisfies the geodesic equation
\begin{equation}
\pddiff{\geo}{\lambda}+\Chr{\alpha}{\beta}{\gamma}
\pdiff{z_G^{\beta}}{\lambda}\pdiff{z_G^{\gamma}}{\lambda} 
= \kappa_G(\lambda)\pdiff{\geo}{\lambda},
\end{equation}
where $\kappa_G(\lambda)$ is the measure of non-affinity of $\lambda$
on the geodesic. Subtracting this geodesic equation from the equation
of motion~\eqref{eq mot} and using Eq.~\eqref{osc 2} to remove the
Christoffel terms, we obtain  
\begin{equation}
\ddiff{\orbit}{\lambda} = \pddiff{\geo}{\lambda}
+ f^{\alpha}\left(\diff{\tau}{\lambda}\right)^2 
+ \left[\kappa(\lambda)
- \kappa_G(\lambda)\right]\pdiff{\geo}{\lambda}. 
\end{equation}
But differentiating Eq.~\eqref{osc 2} yields
$\ddiff{\orbit}{\lambda}=\pddiff{\geo}{\lambda} 
+\left(\pdiff{}{I^A}\pdiff{\geo}{\lambda}\right)\diff{I^A}{\lambda}$.
Comparing these results, we find   
\begin{eqnarray}
\left(\pdiff{}{I^A}\pdiff{\geo}{\lambda}\right)\diff{I^A}{\lambda} 
& = & f^{\alpha}\left(\diff{\tau}{\lambda}\right)^2 
\nonumber\\ 
&& + \left[\kappa(\lambda)-\kappa_G(\lambda)\right]
\pdiff{\geo}{\lambda}. 
\label{diffI 2}
\end{eqnarray}

Equations \eqref{diffI 1} and \eqref{diffI 2} form a closed system
of first-order differential equations for the orbital elements
$I^A$. Two sources of change in the orbital elements are apparent: a 
direct source due to the perturbing force $f^{\alpha}$, and an
indirect source due to the change in the affinity of the
parametrization of the accelerated orbit. Determining this second
effect in practice may be somewhat difficult. However, if we use the
affine parameter $\lambda=\tau$ then the equations simplify to  
\begin{eqnarray}
\pdiff{\geo}{I^A}\dot I^A & = & 0,
\label{diffI 1b} \\
\pdiff{\dot z_G^{\alpha}}{I^A}\dot I^A & = & f^{\alpha} 
\label{diffI 2b}.
\end{eqnarray}
These equations can be easily inverted to solve for the derivatives 
$\dot I^A$, which is done in Sec.~\ref{evolution}. If a non-affine
parameter $\lambda$ is required in a specific application, one may
easily find $\diff{I^A}{\lambda}$ by multiplying the above equations
by $\diff{\tau}{\lambda}$, which will also be done in
Sec.~\ref{evolution}. 


\subsection{Geodesics in Schwarzschild spacetime}
\label{geodesics}

We now focus on the specific case of bound orbits in Schwarzschild
spacetime. The osculating orbits in this case are bound geodesics, 
for which we use the parametrization presented in the text by
Chandrasekhar~\cite{Chandra} and described in detail in
Ref.~\cite{parametrization}. This parametrization is given in 
Schwarzschild coordinates and can be easily derived as follows. 

Because of the spherical symmetry of the Schwarzschild spacetime, we 
can freely set $\theta=\pi/2$. The geodesic equations in a
Schwarzschild spacetime with mass parameter $M$ can be easily solved
for the remaining coordinates to find 
\begin{eqnarray}
\dot t & = & E/F,
\label{tdot}\\
\dot{r}^2 & = & E^2-U_{\rm eff}, 
\label{rdot}\\
\dot\phi & = & \frac{L}{r^2}, 
\label{phidot}
\end{eqnarray}
where $F = 1-2M/r$, $E$ and $L$ are constants equal to energy and
angular momentum per unit mass, respectively, the effective potential
is $U_{\rm eff}=F(1+L/r^2)$, and an overdot represents a derivative
with respect to the proper time $\tau$ on the orbit.  

We are interested in bound orbits that oscillate between a minimal
radius $r_1$ and a maximal radius $r_2$, respectively referred to as
periapsis and apoapsis. Adapting the tradition of celestial mechanics, 
we define the (dimensionless) semi-latus rectum $p$ and the
eccentricity $e$ such that the turning points are given by 
\begin{eqnarray}
r_1 & = & \frac{pM}{1+e}, \label{periapsis}\\
r_2 & = & \frac{pM}{1-e}, \label{apoapsis}
\end{eqnarray}
where $0 \leq e < 1$. These two constants describe the geometry of the 
orbit, just as in Keplerian orbits: $p$ is a measure of the radial
extension of the orbit, while $e$ is a measure of its deviation from
circularity. These constants can be related to $E$ and $L$ by letting
$\dot{r}=0$ in Eq.~\eqref{rdot}, which leads to 
\begin{eqnarray}
E^2 & = & \frac{(p-2-2e)(p-2+2e)}{p(p-3-e^2)}, \label{E}\\
L^2 & = & \frac{p^2M^2}{p-3-e^2}. \label{L}
\end{eqnarray}

Continuing to exploit the analogy with Keplerian orbits, we introduce
a parameter $\chi$ that runs from 0 to $2\pi$ over one radial cycle,
such that $r(\chi)$ takes the elliptical form   
\begin{equation}
r(\chi) = \frac{pM}{1+e\cos(\chi-w)},
\label{r}
\end{equation}
where $w$ is the value of $\chi$ at periapsis, referred to as the
argument of periapsis. The radial component of the velocity is hence 
\begin{equation}
r'(\chi) = \frac{pMe\sin(\chi-w)}
{\bigl[ 1 + e\cos(\chi-w) \bigr]^2},
\label{rprime}
\end{equation}
where a prime henceforth indicates a derivative with respect to 
$\chi$. 

From these results we can relate the parameter $\chi$ to the proper 
time $\tau$ using $\diff{\tau}{\chi}=\frac{r'}{\dot r}$, which yields 
\begin{equation}
\diff{\tau}{\chi} = \frac{p^{3/2}M (p-3-e^2)^{1/2}}
{(p-6-2e\cos v)^{1/2}(1+e\cos v)^2},
\label{chidot}
\end{equation}
where we have introduced the variable 
\begin{equation} 
v \equiv \chi - w
\end{equation} 
for brevity. Along with Eqs.~\eqref{tdot}, \eqref{phidot}, \eqref{E},
and \eqref{L}, this leads to the following parametrizations for
$t(\chi)$ and $\phi(\chi)$: 
\begin{eqnarray}
\phi(\chi) & = & \Phi + \int^{\chi}_w\phi'(\tilde{\chi})d\tilde{\chi}, 
\label{phi}\\
\phi '(\chi) & = & \sqrt{\frac{p}{p-6-2e\cos v}}, 
\label{phiprime}\\
t(\chi) & = & T + \int^{\chi}_wt'(\tilde{\chi})d\tilde{\chi}, 
\label{t}\\
t'(\chi) & = & \frac{p^2M}{(p-2-2e\cos v)(1+e\cos v)^2}
\nonumber\\
&& \times \sqrt{\frac{(p-2-2e)(p-2+2e)}{p-6-2e\cos v}},
\label{tprime}
\end{eqnarray}
where we have defined the constants $T$ and $\Phi$ as the values of 
$t$ and $\phi$ at periapsis, respectively. 

Our parametrization of bound geodesics consists of Eqs.~\eqref{r},
\eqref{rprime}, and \eqref{phi}--\eqref{tprime}. We see that a
geodesic is uniquely specified by the orbital elements
$I^A=\{p,e,w,T,\Phi\}$. The principal elements $p$ and $e$
determine the spatial shape of the orbit and are equivalent to
specifications of energy and angular momentum; they determine the
choice of geodesic. The positional elements $w$, $T$, and
$\Phi$ determine the spatial orientation and time-dependence of the
orbit; they determine the starting point of the particle on the
selected geodesic. All together, the specification of the orbital
elements is equivalent to the specification of initial values for the
position and velocity of the particle. We need three initial positions  
for a planar orbit, and we need two initial velocities (three minus
one, by virtue of the normalization condition on the velocity
vector); this counting matches the number of orbital elements.  

%
We note that our choice of orbital elements is closely related to
Mino's in Ref.~\cite{Mino2005}. When the orbital motion is restricted 
to the equatorial plane of a Kerr black hole, Mino uses the
principal elements $E$ and $L$ and positional elements that are
identical to our $w$, $T$, and $\Phi$. To use $(p,e)$ instead of
$(E,L)$ is mostly a matter of taste; we believe that the set $(p,e)$
is more useful than $(E,L)$ because it gives a simpler
parametrization, and because $p$ and $e$ are geometrically more  
informative. In the following subsection we will deviate more strongly  
from Mino's parametrization: for reasons that will be explained, we
shall avoid directly evolving the elements $T$ and $\Phi$.  
%

All the equations presented in this section remain valid for a
perturbed orbit, with the exception of Eqs.~\eqref{periapsis} and
\eqref{apoapsis}, which lose their meaning. The alteration that we
shall make to account for the perturbation is that in each equation,
the orbital elements will become functions of $\chi$. 


\subsection{Evolution equations}
\label{evolution}

If we restrict the perturbing force to lie in the plane of the orbit, 
and assume that the orbit remains bound, then the geodesics described
in the last section form a sufficient set of osculating orbits. Using
our parametrization of these geodesics, along with the results of our 
general analysis in Sec.~\ref{general case}, we can now find evolution 
equations for the orbital elements. Multiplying both sides of
Eq.~\eqref{diffI 1b} by $\diff{\tau}{\chi}$, we find 
\begin{eqnarray}
\pdiff{r}{p}p' + \pdiff{r}{e}e' + \pdiff{r}{w}w'  & = & 0, 
\label{wdot}\\
\pdiff{t}{p}p' + \pdiff{t}{e}e' + \pdiff{t}{w}w'  + T' & = & 0, 
\label{Tdot}\\
\pdiff{\phi}{p}p' + \pdiff{\phi}{e}e' + \pdiff{\phi}{w}w' 
+ \Phi' & = & 0.
\label{Phidot} 
\end{eqnarray}
Similarly, from Eq.~\eqref{diffI 2b} we find
\begin{eqnarray}
\pdiff{\dot t}{p}p' + \pdiff{\dot t}{e}e' + \pdiff{\dot t}{w}w' 
& = & f^t\tau', 
\label{ft}\\
\pdiff{\dot r}{p}p' + \pdiff{\dot r}{e}e' + \pdiff{\dot r}{w}w' 
& = & f^r\tau', 
\label{fr}\\
\pdiff{\dot \phi}{p}p' + \pdiff{\dot \phi}{e}e' 
+ \pdiff{\dot\phi}{w}w' & = & f^{\phi}\tau'. 
\label{fphi}
\end{eqnarray}

%
The orthogonality condition $f^{\alpha}\dot z_{\alpha}=0$ allows us to
remove one component of Eq.~\eqref{diffI 2b} from the set of
equations; we use this freedom to remove Eq.~\eqref{ft}. The remaining 
equations decouple into a closed system of ordinary differential
equations for $p$, $e$, and $w$ and two auxiliary equations for $T$
and $\Phi$. We shall find that the evolution equations for $p$, $e$,
and $w$ are simple. The equations for $T$ and $\Phi$, however, are
not: Factors such as $\pdiff{t}{p}$ in Eqs.~\eqref{Tdot} and
\eqref{Phidot} introduce elliptic integrals of the form 
$\int_w^\chi\pdiff{t'}{p}(\tilde\chi)d\tilde\chi$ into the expressions
for $T'$ and $\Phi'$. These integrals would have to be evaluated at 
each time-step in a numerical evolution, and they would create an
excessive computational cost. Additionally, the integrals generally
grow linearly with $\chi$, and this produces terms in $T(\chi)$ and
$\Phi(\chi)$ that grow quadratically with $\chi$, as well as terms
that oscillate with a linearly increasing amplitude. Such terms
greatly confuse both numerical and analytical descriptions, and they
are largely an artefact of our parametrization. (This statement applies
also to Mino's parametrization \cite{Mino2005}.) We note that
similar (though less severe) difficulties arise also in the method of 
osculating orbits in Newtonian celestial mechanics; refer for example
to the discussion on pp. 248--250 in the text by Beutler
\cite{Beutler}. In the Newtonian context, alternative orbital elements 
are typically selected so as to overcome these problems. With no
obvious choice of alternative elements in the relativistic context, we
opt instead to directly evolve the coordinates $t$ and $\phi$ rather
than the elements $T$ and $\Phi$.     
%

Our phase space thus consists of $\{p,e,w,t,\phi\}$. This choice of 
phase space does not allow an easy separation of perturbative from
geodesic effects in the evolutions of $t$ and $\phi$, nor does
it allow a clean separation of conservative from dissipative effects. 
But it is overwhelmingly more convenient than the alternative choice 
$\{p,e,w,T,\Phi\}$. If $T$ and $\Phi$ are required in an application, 
they may be found as, e.g., $T=t-\int_w^\chi
t'(\tilde{\chi})d\tilde{\chi}$. This may be necessary if initial 
conditions are required on an osculating orbit, or if one wishes to
fully isolate perturbative effects. 

Solving for $w'$ from Eq.~\eqref{wdot}, and noting that
$\pdiff{r}{w}=-r'$, we find 
\begin{equation}
w' =\frac{1}{r'}\left(\pdiff{r}{p}p' + \pdiff{r}{e}e'\right).
\label{wprime_raw} 
\end{equation}
Substituting this into Eqs.~\eqref{ft} and \eqref{fphi}, we can solve
for $p'$ and $e'$ to find 
\begin{eqnarray}
p' & = & \frac{\mathcal{L}_e(\phi)f^r-\mathcal{L}_e(r)f^{\phi}}
{\mathcal{L}_e(\phi)\mathcal{L}_p(r)-\mathcal{L}_e(r)\mathcal{L}_p(\phi)}
\tau', \\
e' & = & \frac{\mathcal{L}_p(r)f^{\phi}-\mathcal{L}_p(\phi)f^r}
{\mathcal{L}_e(\phi)\mathcal{L}_p(r)-\mathcal{L}_e(r)\mathcal{L}_p(\phi)}
\tau',
\end{eqnarray}
where $\mathcal{L}_a(x)\equiv \pdiff{\dot x}{a}
+ \frac{1}{r'}\pdiff{r}{a}\pdiff{\dot x}{w}$. Explicitly, the
results are 
\begin{widetext}
\begin{eqnarray}
p' & = & \frac{2p^{7/2}M^2(p-3-e^2)(p-6-2e\cos v)^{1/2} 
  (p-3-e^2\cos^2v)}{(p-6+2e)(p-6-2e)(1+e\cos v)^4} f^{\phi}
- \frac{2p^3Me(p-3-e^2)\sin v}{(p-6+2e)(p-6-2e)(1+e\cos v)^2} f^r,
\quad  
\label{pprime} \\
e' & = & \frac{p^{5/2}M^2(p-3-e^2)\left\lbrace(p-6-2e^2)
  \left[(p-6-2e\cos v)e\cos v+2(p-3)\right]\cos v+e(p^2-10p+12+4e^2)
  \right\rbrace}{(p-6+2e)(p-6-2e)(p-6-2e\cos v)^{1/2}
  (1+e\cos v)^4} f^{\phi} 
\nonumber\\
&& + \frac{p^2M(p-3-e^2)(p-6-2e^2)\sin v}{(p-6+2e)
  (p-6-2e)(1+e\cos v)^2} f^r, 
\label{eprime} \\
w' & = & \frac{p^{5/2}M^2(p-3-e^2)\left\lbrace(p-6)
\left[(p-6-2e\cos v)e\cos v+2(p-3)\right]-4e^3\cos v\right\rbrace 
  \sin v}{e(p-6+2e)(p-6-2e)(p-6-2e\cos v)^{1/2}(1+e\cos v)^4} f^{\phi} 
\nonumber\\
&& -\frac{p^2M(p-3-e^2)\left[(p-6)\cos v+2e\right]}{e(p-6+2e)(p-6-2e)
  (1+e\cos v)^2} f^r.
\label{wprime}
\end{eqnarray}
\end{widetext}
These equations could be rewritten in any number of ways, in terms of
alternative linear combinations of $f^t$, $f^r$, and $f^{\phi}$, by
using the orthogonality relation $f_{\alpha}\dot{z}^{\alpha}=0$, which
has the explicit form  
\begin{equation}\label{orthogonality}
F t' f^t - F^{-1} r' f^r - r^2 \phi' f^{\phi} = 0. 
\end{equation}
The result of such a rearrangement might in fact be simpler, but it
may also be ill-behaved from a numerical point of view. One such
alternative combination is given in 
Appendix~\ref{elements from E and L}.  

Our first formulation of the method of osculating orbits is
complete. We have first-order evolution equations for each one of the
dynamical variables in the set $\{p,e,w,t,\phi\}$; the equations for
$t$ and $\phi$ were obtained in the preceding subsection, and for
convenience they are reproduced here: 
\begin{eqnarray} 
t' & = & \frac{p^2M}{(p-2-2e\cos v)(1+e\cos v)^2}
\nonumber\\
&& \times \sqrt{\frac{(p-2-2e)(p-2+2e)}{p-6-2e\cos v}}, 
\\ 
\phi' & = & \sqrt{\frac{p}{p-6-2e\cos v}}. 
\end{eqnarray}
Equations (\ref{pprime}), (\ref{eprime}), and (\ref{wprime}) form a
complete set of equations for $p(\chi)$, $e(\chi)$, and $w(\chi)$;
once these functions are known, $t(\chi)$ and $\phi(\chi)$ can be
obtained from the remaining two equations. We recall that $v = \chi -
w(\chi)$.   

One may note that $w'$ diverges as $e \to 0$. This corresponds to the 
fact that $w$ loses its geometric meaning for circular orbits. To
overcome this difficulty we can again follow celestial mechanics and 
define alternative orbital elements $\alpha=e\sin w$ and $\beta=e\cos
w$. The radial coordinate in terms of these elements is  
\begin{equation}
r=\frac{pM}{1+\Psi+\Omega},
\end{equation}
where $\Psi=\alpha\sin\chi$ and $\Omega=\beta\cos\chi$ are introduced
for the sake of brevity in later expressions. While $\alpha$ and
$\beta$ do not possess a clear geometric meaning, which limits their 
usefulness for generic orbits, they do allow one to analyze
small-eccentricity or quasi-circular orbits. Their evolution equations
can be easily calculated as $\alpha '=e'\sin w+ew'\cos w$ and $\beta
'=e'\cos w-ew'\sin w$. Using the identities $e\cos
v=\alpha\sin\chi+\beta\cos\chi$ and $e\sin
v=\beta\sin\chi-\alpha\cos\chi$ to simplify the results, we find 
\begin{widetext}
\begin{eqnarray}
\beta ' & = & \frac{p^{5/2}M^2(p-3-\alpha^2-\beta^2)f^{\phi}}
{\sqrt{p-6-2(\Psi+\Omega)}((p-6)^2-4(\alpha^2+\beta^2))(1+\Psi+\Omega)^4}   
\times\Bigg\lbrace-4\alpha\left[\alpha\beta\cos2\chi
+\frac{1}{2}(\alpha^2\!-\!\beta^2)\sin2\chi\right]\nonumber\\ 
&& +\left[2(p-3)+(p-6)(\Psi\!+\!\Omega)-2(\Psi\!+\!\Omega)^2\right]
\left[(p-6)\cos\chi-2\beta(\Psi\!+\!\Omega)\right]
+\beta\left[p^2-10p+12+4(\alpha^2\!+\!\beta^2)\right] 
\!\!\!\Bigg\rbrace \nonumber\\
&& +\frac{p^2M(p-3-\alpha^2-\beta^2)
\left[(p-6-2\beta^2)\sin\chi+2\alpha(1+\Omega)\right]f^r} 
{((p-6)^2-4(\alpha^2+\beta^2))(1+\Psi+\Omega)^2},\\
\alpha ' & = & \frac{p^{5/2}M^2(p-3-\alpha^2-\beta^2)f^{\phi}}
{\sqrt{p-6-2(\Psi+\Omega)}((p-6)^2-4(\alpha^2+\beta^2))(1+\Psi+\Omega)^4}  
\times\Bigg\lbrace 
4\beta\left[\alpha\beta\cos2\chi+\frac{1}{2}(\alpha^2\!
- \!\beta^2)\sin2\chi\right]\nonumber\\
&& +\left[2(p-3)+(p-6)(\Psi\!+\!\Omega)-2(\Psi\!+\!\Omega)^2\right]
\left[(p-6)\sin\chi-2\alpha(\Psi\!+\!\Omega)\right]
+\alpha\left[p^2-10p+12+4(\alpha^2\!+\!\beta^2)\right]
\!\!\!\Bigg\rbrace \nonumber\\
&& -\frac{p^2M(p-3-\alpha^2-\beta^2)
\left[(p-6-2\alpha^2)\cos\chi+2\beta(1+\Psi)\right]f^r} 
{((p-6)^2-4(\alpha^2+\beta^2))(1+\Psi+\Omega)^2}.
\end{eqnarray}
To evolve our full system we must also express $p'$, $t'$, and 
$\phi'$ in terms of $\alpha$ and $\beta$: 
\begin{eqnarray}
p' & = & \frac{2p^{7/2}M^2\sqrt{p-6-2(\Psi+\Omega)}
(p-3-\alpha^2-\beta^2)(p-3-(\Psi+\Omega)^2)f^{\phi}}
{[(p-6)^2-4(\alpha^2+\beta^2)](1+\Psi+\Omega)^4}\nonumber\\ 
&&
-\frac{2p^3M(p-3-\alpha^2-\beta^2)(\beta\sin\chi-\alpha\cos\chi)f^r}
{[(p-6)^2-4(\alpha^2+\beta^2)](1+\Psi+\Omega)^2},\\ 
\nonumber\\
t'(\chi) & = &
\frac{p^2M\sqrt{(p-2)^2-4(\alpha^2+\beta^2)}}
{(p-2-2(\Psi+\Omega))\sqrt{p-6-2(\Psi+\Omega)}(1+\Psi+\Omega)^2},\\ 
\nonumber\\\phi '(\chi) & = & \sqrt{\frac{p}{p-6-2(\Psi+\Omega)}}.
\end{eqnarray}
\end{widetext}
This is our second formulation of the method of osculating orbits. The
first formulation involves shorter equations, but it becomes
ill-behaved when $e$ is small. The second formulation is well
behaved, but it involves longer equations. 


\section{Post-Newtonian binaries}
\label{self-force}

\subsection{Hybrid equations of motion}
\label{hybrid}

We now move on to a concrete application of our method by considering 
the post-Newtonian binary system introduced in 
Sec.~\ref{PN binaries}. This system consists of two
gravitationally-bound bodies of mass $m_1$ and $m_2$, with equations
of motion derived to 2.5PN order in a post-Newtonian expansion;
because we are interested in self-force effects, we take the ratio
$m_1/m_2$ to be small, and we neglect the spin of the bodies. In
this section we explain how such a system can be analyzed with our
method of osculating orbits.  

Our analysis is based upon the hybrid equations of motion presented in
Ref.~\cite{Kidder}. These equations begin with the 2.5PN equations of
motion for each one of the two bodies. Within the center-of-mass frame
the relative motion of the two bodies is governed by the closed system
of equations \cite{Lincoln}  
\begin{equation}\label{PNeqns}
\ddiff{x^a_h}{t} = -\frac{M}{r_h^2}\left(A\frac{x_h^a}{r_h}
+ B\diff{x_h^a}{t}\right),
\end{equation}
where $x_h^a\equiv x^a_1-x^a_2$ is a Cartesian spatial vector from 
$m_2$ to $m_1$ in harmonic coordinates, $r_h^2=\delta_{ab}x^ax^b$ is
the square of the vector's Euclidean magnitude, $t$ is a harmonic time 
coordinate, and $M = m_1+m_2$ is the total mass of the system. The
functions $A$ and $B$ depend only on the total mass $M$, the reduced
mass $\mu=m_1m_2/M$, and the relative coordinates and
velocities. They can be written as $A=A_M+\epsilon\tilde{A}$ and
$B=B_M+\epsilon\tilde{B}$, where $\epsilon=\mu/M$ and terms with a
subscript $M$ are independent of $\mu$. The $\mu$-dependent terms are
quadratic in $\epsilon$, and they can be further decomposed into
post-Newtonian orders as $\tilde{A} = \tilde{A}_1 + \tilde{A}_2 
+ \tilde{A}_{2.5}$ and $\tilde{B} = \tilde{B}_1 + \tilde{B}_2 
+ \tilde{B}_{2.5}$. Explicitly, these have the form 
\begin{eqnarray}
A_M & = &
1-4\frac{M}{r_h}+v^2+9\left(\frac{M}{r_h}\right)^2-2\frac{M}{r_h}
\left(\diff{r_h}{t}\right)^2,\\ 
\epsilon\tilde{A_1} & = & -\epsilon\left[2\frac{M}{r_h}-3v^2
+\frac{3}{2}\left(\diff{r_h}{t}\right)^2\right], \\
\epsilon\tilde{A_2} & = &
\epsilon\bigg[\frac{87}{4}\left(\frac{M}{r_h}\right)^2 
+ (3-4\epsilon)v^4-\frac{1}{2}(13-4\epsilon) 
\frac{M}{r_h}v^2\nonumber\\
 && \phantom{\epsilon\bigg[}-\frac{3}{2}(3-4\epsilon)v^2
\left(\diff{r_h}{t}\right)^2+\frac{15}{8}(1-3\epsilon)
\left(\diff{r_h}{t}\right)^4 \nonumber\\ 
&& \phantom{\epsilon\bigg[} -(25+2\epsilon)\frac{M}{r_h}
\left(\diff{r_h}{t}\right)^2\bigg],
\end{eqnarray}
\begin{eqnarray}
\epsilon\tilde A_{2.5} & = & -\frac{8}{5}\epsilon\frac{M}{r_h}
\diff{r_h}{t}\left[3v^2+\frac{17}{3}\frac{M}{r_h}\right],\\
\nonumber\\
B_M & = &  -\diff{r_h}{t}\left(4-2\frac{M}{r_h}\right),\\ 
\epsilon\tilde{B}_1 & = & 2\epsilon\diff{r_h}{t},\\
\epsilon\tilde{B}_2 & = & -\frac{1}{2}\epsilon\diff{r_h}{t}
\bigg[(15+4\epsilon)v^2-(41+8\epsilon)\frac{M}{r_h} \nonumber\\ 
&& \phantom{-\frac{1}{2}\epsilon\diff{r_h}{t}\bigg[}
    -3(3+2\epsilon)\left(\diff{r_h}{t}\right)^2\bigg],\\ 
\epsilon\tilde{B}_{2.5} & = &
\frac{8}{5}\epsilon\frac{M}{r_h}\left[v^2+3\frac{M}{r_h}\right], 
\end{eqnarray}
where
$v^2 \equiv \delta_{ab} \diff{x^a_h}{t} \diff{x^b_h}{t}$ is the
square of the velocity vector in harmonic coordinates.   
 
The hybrid equations are inspired by the fact that when $\epsilon=0$, 
Eq.~\eqref{PNeqns} becomes identical to a 2PN expansion of the
geodesic equation in a Schwarzschild spacetime with mass parameter
$M$. Building on this fact, Kidder, Will, and Wiseman \cite{Kidder}
replaced $A_M$ and $B_M$ with their exact geodesic expressions $A_S$
and $B_S$ in the fictitious Schwarzschild spacetime. In other words,
the hybrid equations of motion are given by Eq.~\eqref{PNeqns} after
substituting $A=A_S+\epsilon\tilde{A}$ and $B=B_S+\epsilon\tilde{B}$,
where  
\begin{eqnarray}
A_S & = & \frac{1-M/r_h}{(1+M/r_h)^3} \nonumber\\ 
&&-\frac{2-M/r_h}{1-M^2/r_h^2}\frac{M}{r_h}
\left(\diff{r_h}{t}\right)^2+v^2, \\
B_S &=&-\frac{4-2M/r_h}{1-M^2/r_h^2}\diff{r_h}{t}.
\end{eqnarray}
The resulting equations are accurate to 2.5PN order, but in the
test-mass limit $m_1\to 0$ they exactly describe the orbit of the test
mass in the Schwarzschild spacetime of the other body. These equations
form an ideal test case for our method of osculating orbits because,
besides their relative simplicity, they explicitly split into geodesic
terms and perturbation terms. This allows us to construct osculating
orbits as geodesics in the fictitious Schwarzschild spacetime of mass
$M$. We can then easily derive the perturbing force from the terms
$\tilde{A}$ and $\tilde{B}$. 

The first step in this process is to write the equations of motion in
plane polar coordinates $(r_h,\phi)$, which are defined by
$x^1_h=r_h\cos(\phi)$ and $x^2_h=r_h\sin(\phi)$. In terms of these
coordinates, Eq.~\eqref{PNeqns} becomes 
\begin{eqnarray}
\ddiff{r_h}{t} & = & -\frac{M}{r_h^2}\left(A+B\diff{r_h}{t}\right)
+ r_h\left(\diff{\phi}{t}\right)^2,\label{r acceleration}\\ 
\ddiff{\phi}{t} & = & -\frac{M}{r_h^2}B\diff{\phi}{t}
- \frac{2}{r_h}\diff{r_h}{t}\diff{\phi}{t}.\label{phi acceleration}
\end{eqnarray}
The harmonic coordinates used here are related to Schwarzschild
coordinates by the simple transformation $r_h = r-M$. Since $M$ is
constant, the subscript $h$ can be safely dropped within 
derivatives. Expressing $r_h$ in terms of $r$, the above equations are 
transformed into Schwarzschild coordinates. 

We derive $f^{\alpha}$ from these equations as follows. From
Eq.~\eqref{eq mot} we have 
\begin{equation}
f^{\alpha} = \dot t^2\left(\ddiff{\orbit}{t}
+\Chr{\alpha}{\beta}{\gamma}\diff{z^{\beta}}{t}\diff{z^{\gamma}}{t}
-\kappa(t) \diff{\orbit}{t}\right).\label{force 1}
\end{equation}
Although we could calculate $\kappa(t)$ directly from its definition,
the result would be unwieldy. We instead use the equation of motion
for $t$, 
\begin{equation}
\ddiff{t}{t}+\Chr{t}{\beta}{\gamma}\diff{z^{\beta}}{t}\diff{z^{\gamma}}{t}  
= f^t\dot t^{-2} + \kappa\diff{t}{t}, 
\end{equation}
to replace $\kappa$ with
\begin{equation}
\kappa = \Chr{t}{\beta}{\gamma}\diff{z^{\beta}}{t}\diff{z^{\gamma}}{t}
- f^t\dot t^{-2}.
\end{equation}
Substituting this expression for $\kappa$ into Eq.~\eqref{force 1}, we
find  
\begin{equation}
f^{\alpha} = \dot t^2 a^\alpha_p + \diff{\orbit}{t}f^t, 
\label{force from d^2xdt^2} 
\end{equation}
where
\begin{equation}
a^\alpha_p \equiv
\ddiff{\orbit}{t}+\Big(\Chr{\alpha}{\beta}{\gamma} -
\diff{\orbit}{t}\Chr{t}{\beta}{\gamma}\Big)
\diff{z^{\beta}}{t}\diff{z^{\gamma}}{t}. 
\end{equation}
The subscript $p$ refers to the fact that $a^\alpha_p$ involves only
the perturbative terms in $d^2\orbit/dt^2$. Indeed, a
simple calculation based on the preceding equations for $d^2r/dt^2$
and $d^2\phi/dt^2$, as well as the Christoffel symbols obtained from
the Schwarzschild metric, reveals that 
\begin{eqnarray}
a^r_p &=& -\frac{M}{r_h^2}\left(\epsilon\tilde A
+\epsilon\tilde B\diff{r}{t}\right), \\
a^\phi_p &=& -\frac{M}{r_h^2}\epsilon\tilde B\diff{\phi}{t}. 
\end{eqnarray}

Equation~\eqref{force from d^2xdt^2} determines $f^r$ and $f^{\phi}$ 
in terms of $f^t$. The orthogonality condition
\eqref{orthogonality} then allows us to find all three components of
the force. The result is 
\begin{eqnarray}
f^t\!&=&\!\displaystyle\frac{\dot t^2\!
\left[a^r_p \diff{r}{t} + a^\phi_p r^2F \diff{\phi}{t}\right]}
{F^2 - \big(\diff{r}{t}\big)^{2} 
- Fr^2\big(\diff{\phi}{t}\big)^{2}},\\
f^r\! &=& \!\frac{\dot t^2\!
\left[a^r_p \Big(F - r^2\big(\diff{\phi}{t}\big)^{2}\Big) 
+ a^\phi_p r^2\diff{r}{t}\diff{\phi}{t}\right]}
  {F^{-1}\Big(F^2 - \big(\diff{r}{t}\big)^{\!2} 
- Fr^2\big(\diff{\phi}{t}\big)^{2}\Big)},\\
f^{\phi}\! &=& \!\frac{\dot t^2\!
\left[a^r_p \diff{r}{t}\diff{\phi}{t} 
+ a^\phi_p \Big(F^2 - \big(\diff{r}{t}\big)^2\Big)\right]}
  {F^2 - \big(\diff{r}{t}\big)^{2} 
- Fr^2\big(\diff{\phi}{t}\big)^{2}}. 
\end{eqnarray}
Substituting $a^\alpha_p$ into the above results, and using the
normalization condition $-1=\dot z^{\alpha} \dot z_{\alpha} 
=-F\dot t^2 + F^{-1}\dot r^2 +r^2 \dot\phi^2$, leads to 
\begin{eqnarray}
f^r & = & -\frac{\epsilon M \dot{t}^4}{r_h^2} \Bigl\{ 
\bigl[ F - r^2 (d\phi/dt)^2 \bigr] \tilde{A} 
+ F (dr/dt) \tilde{B} \Bigr\}, 
\nonumber \\ 
f^\phi & = & -\frac{\epsilon M \dot{t}^4}{r_h^2} \frac{d\phi}{dt} 
\Bigl\{ F^{-1} (dr/dt) \tilde{A} + F \tilde{B} \Bigr\}.
\end{eqnarray} 
Since $f^t$ is not required in our formalism, we will not provide an 
explicit expression for it.  

We can recast these equations in a form analogous to that of
Eqs.~\eqref{r acceleration} and \eqref{phi acceleration},  
\begin{eqnarray}
f^r & = & -\frac{\mu}{r^2} 
\left[\mathcal{A} + \mathcal{B} \frac{dr}{dt} \right],
\\ 
f^{\phi} & = &-\frac{\mu}{r^2}\mathcal{B} \frac{d\phi}{dt},  
\end{eqnarray}
by defining $\mathcal{A}$ and $\mathcal{B}$ as
\begin{eqnarray}
\mathcal{A} & = & \frac{\dot{t}^2}{(1-M/r)^2} \tilde{A}, 
\label{script A} \\ 
\mathcal{B} & = & \frac{\dot{t}^4}{(1-M/r)^2}
\left( \frac{1}{F} \frac{dr}{dt} \tilde{A} + F \tilde{B} \right). 
\label{script B} 
\end{eqnarray}
The factors of $\dot t$ convert the ``time" variable in the
acceleration from coordinate time to proper time; this is given by 
\begin{equation} 
\dot{t}^2 = \frac{1}{F - F^{-1} (dr/dt)^2 - r^2 (d\phi/dt)^2}, 
\end{equation}
where, we recall, $F = 1-2M/r$. The factors of $1/(1-M/r)^2$, on the
other hand, convert from harmonic coordinates to Schwarzschild
coordinates. One could incorporate these factors into each 
$\tilde A_i$ and $\tilde B_i$ and then re-expand these in powers
of $M/r$ to find new expressions for $\mathcal{A}_i$ and
$\mathcal{B}_i$, neglecting terms of 3PN order and higher; but since
the hybrid equations already introduce errors above 2.5PN order, doing
so is unnecessary. Thus, for simplicity we shall use the force in its
above form.  

The final expression for the perturbing force is obtained by
substituting the post-Newtonian expansions for $\tilde{A}$ and
$\tilde{B}$ into Eqs.~(\ref{script A}) and (\ref{script B}); the
relevant equations are listed near the beginning of
Sec.~\ref{hybrid}. In these equations we must make the substitution
$r_h = r-M$, and convert $t$-derivatives into $\chi$-derivatives by
employing Eq.~(\ref{tprime}). In these final forms, the expressions
for $f^r$ and $f^\phi$ are ready to be inserted within the evolution
equations for the orbital elements. 


\subsection{Results}
\label{results}

\subsubsection{Adiabatic, secular, and radiative approximations}

%
We are primarily interested in determining the types of errors
introduced by the adiabatic and radiative approximations. We should
first clarify the meaning of these approximations. The basis of both
approximations in the context of osculating orbits is the separation
of orbital elements into secular and oscillating parts, i.e. 
$I^A=I^A_{\mathrm{sec}} + I^A_{\mathrm{osc}}$. The particular
adiabatic approximation that we are concerned with, which we have
titled ``secular approximation,'' is one which eliminates the
oscillations and keeps only the secular behavior; that is, it uses an 
approximate orbital evolution with 
$I^A_{\mathrm{adb}}=I^A_{\mathrm{sec}}$. A radiative approximation 
uses only dissipative terms in the perturbing force, with orbital
elements $I^A_{\mathrm{r}}$, with the hope that the secular part 
$I^A_{\mathrm{r\ sec}}$ of this evolution reproduces
$I^A_{\mathrm{sec}}$.
%

\begin{figure}[tb]
\begin{center}
\includegraphics{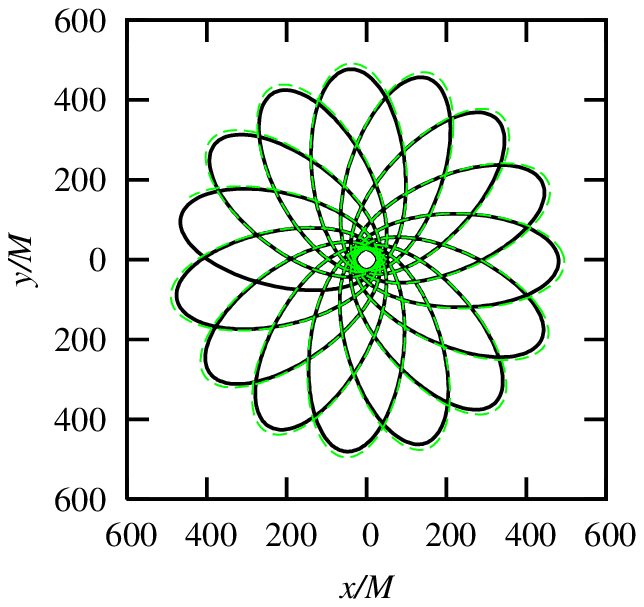}
\includegraphics{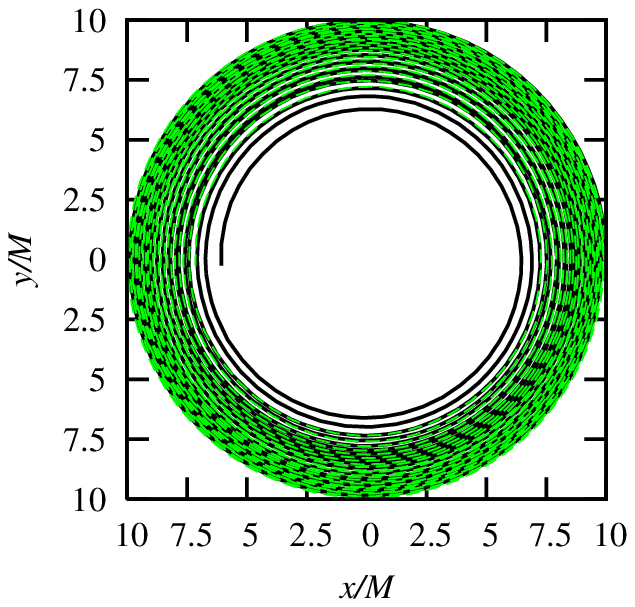}
\end{center}
\caption{Comparisons of true orbits (solid black curves) and radiative
  approximation orbits (dashed green curve) with identical initial
  conditions and with a mass ratio $\mu/M=0.01$. In each case the two
  orbits begin at periapsis and are terminated at the same final
  time. Upper plot: highly eccentric orbits with $p_0=50$ and
  $e_0=0.9$. At the end of the simulation the approximate orbit lags
  behind the true orbit by approximately one-half radial cycle out of
  a total of fifteen. Lower plot: quasi-circular orbits with identical
  initial conditions $p_0=10$ and $e_0=0$. Again, the approximate
  orbit lags behind the true orbit.} 
\label{orbits}
\end{figure}

\begin{figure}[h!]
\begin{center}
\includegraphics{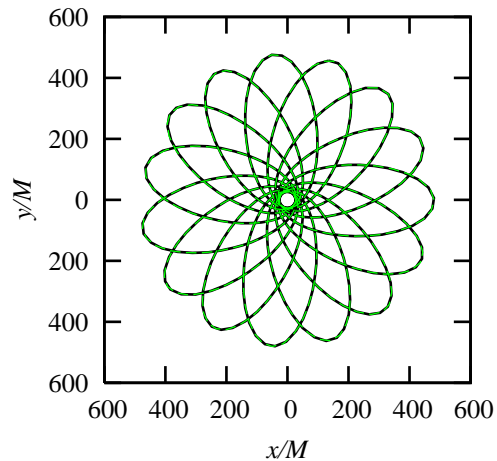}
\end{center}
\caption{The same eccentric orbit as shown in Fig.~\ref{orbits}, but
  now using time-averaged initial conditions for the radiative
  approximation. In this case the approximate orbit is
  indistinguishable from the true orbit on the timescale of the plot
  (fifteen orbital cycles).}  
\label{av_orbit}
\end{figure}

Unfortunately, these general definitions are somewhat ambiguous. We
examine first the case of the secular approximation. The main source
of ambiguity associated with the general idea of removing oscillations
is that it is not clear {\it which oscillations} are intended to be
removed. For example, in the formalism presented in this paper,
removing the oscillations with respect to $\chi$ will not remove the
oscillations with respect to $t$, and vice versa. This failure is
caused by the zeroth-order (i.e. geodesic) oscillations in time as a 
function of $\chi$. Consequently, a secular evolution defined by an 
average over the orbital parameter $\chi$, such as 
\begin{equation}\label{chiav}
I^A_{\mathrm{sec}} =
\langle I^A\rangle_{\chi} \equiv 
\frac{1}{2\pi}\int^{\chi + \pi}_{\chi - \pi} I^A(\chi')\, d\chi',
\end{equation}
will differ from that defined by an average over time, such as
\begin{equation}\label{tav}
I^A_{\mathrm{sec}} = \langle I^A \rangle_t \equiv 
\frac{\int^{\chi + \pi}_{\chi - \pi} I^A \diff{t}{\chi} d\chi'} 
{\int^{\chi + \pi}_{\chi - \pi} \diff{t}{\chi} d\chi'}. 
\end{equation}
A precise definition of a secular approximation would have to
specify which averaging procedure is to be selected. 

\begin{figure}[h!]
\begin{center}
\includegraphics{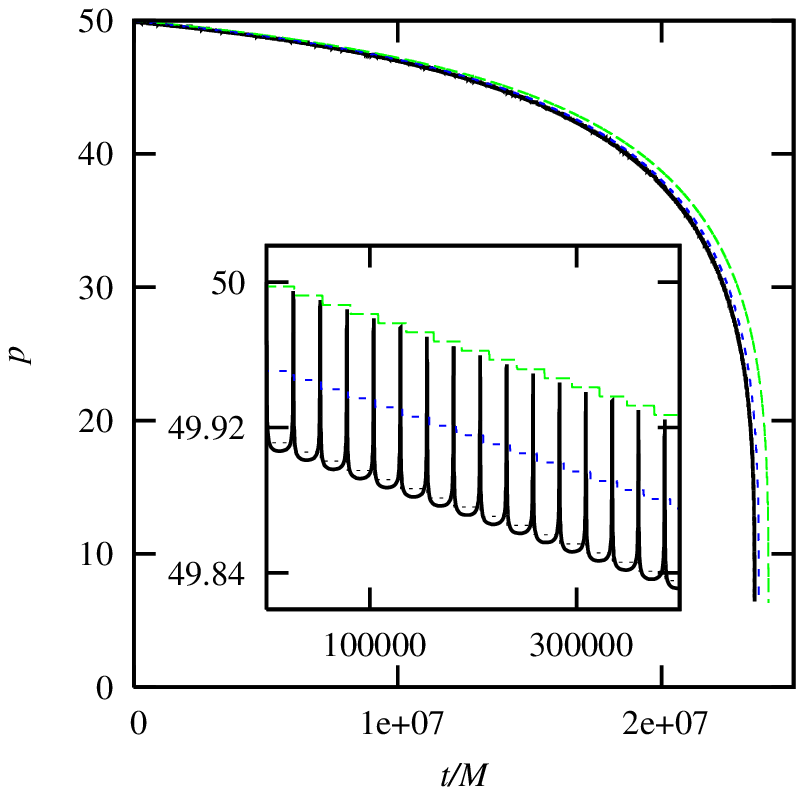}
\includegraphics{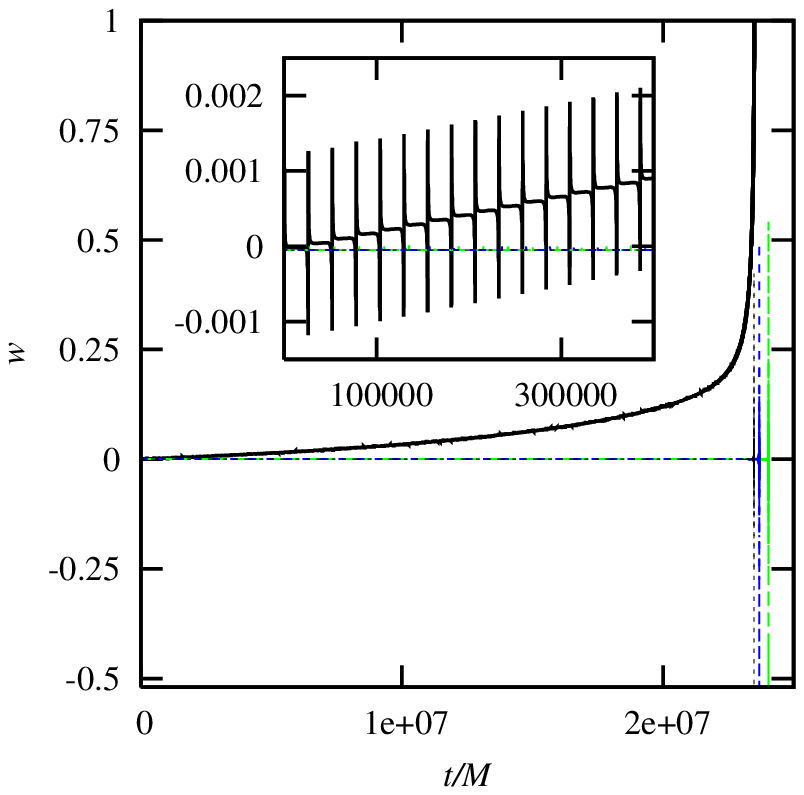}
\end{center}
\caption{The principal element $p$ and positional element $w$ as
  functions of time for a complete inspiral, beginning with the
  initial conditions of the eccentric orbit in Fig.~\ref{orbits}. In
  each plot the true curve is in solid black, the radiative curve with
  the same initial conditions is long-dashed in green (the uppermost
  curve in the $p$ plot), the radiative curve with $\chi$-averaged
  initial conditions is short-dashed in blue (middle curve in $p$
  plot), and the radiative curve with time-averaged initial conditions
  is dotted in black (lowest curve in $p$ plot). The insets display
  the early behavior of the curves, covering the same range of time as
  in Fig.~\ref{orbits}.} 
\label{plots_vs_t}
\end{figure}

A second source of ambiguity concerns the choice of initial
conditions. We desire that our secular evolution reproduce the   
average of the true evolution, and this means that in general, the
initial conditions placed on the approximate solution will have to
differ from the exact initial conditions. This is because the exact 
solution contains the secular approximation plus oscillations, and
the oscillations may not vanish at the initial time. Identifying the
correct initial conditions for the approximate evolution therefore
requires knowledge of the oscillations; in the absence of such
information---that is, when the exact solution is not known---the
initial conditions remain unknown and the procedure is ambiguous. The
ambiguity persists even when the exact solution is known, because it
is then inherited from the first source of ambiguity, the question as
to which oscillations are to be removed. The ambiguity associated with 
the initial conditions is lifted only when the averaging procedure is
selected, and when the exact solution is known; the approximate
initial conditions are then calculated by averaging the exact
evolution over the first radial cycle. For example, in the case of an
averaging over $\chi$ we would set $I^A_{\rm sec}(0) = (2\pi)^{-1} 
\int_0^{2\pi} I^A(\chi)\, d\chi$.  

\begin{figure}[tb]
\begin{center}
\includegraphics{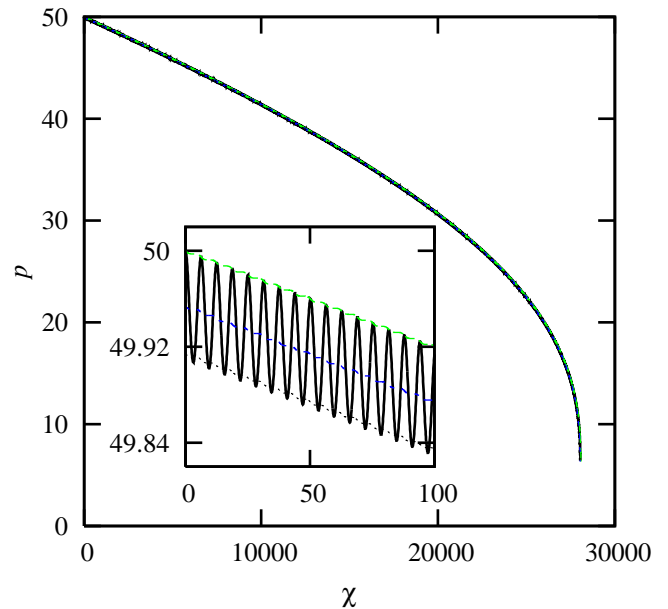}
\end{center}
\caption{The principal element $p$ as a function of the orbital
  parameter $\chi$. The curves are as described in
  Fig.~\ref{plots_vs_t}. The radiative curves do deviate secularly
  from the true curve, but the errors are too small too appear on the
  scale of the graph.} 
\label{plots_vs_chi}
\end{figure}

We shall not pursue a detailed exploration of the ambiguities
associated with the secular approximation in this paper, although they
are quite important; they are the focus of a companion paper
\cite{other_paper}. Our focus here will be instead on the limitations
and ambiguities of the radiative approximation. As was indicated
previously, a radiative evolution switches off all conservative terms 
in the perturbing force ($\tilde{A}_1 = \tilde{A}_2 = \tilde{B}_1 =
\tilde{B}_2 = 0$), and retains only the radiative terms at 2.5PN
order ($\tilde{A}_{2.5} \neq 0$ and $\tilde{B}_{2.5} \neq 0$). This
approximation is logically distinct from adiabatic approximations in
general, but the hope formulated in the literature (for example in 
Refs.~\cite{Mino, Nakano, Hughes}) is that the radiative evolution
will reproduce the secular changes of the orbital elements. We shall
see that while the radiative approximation captures the secular
changes in $p(\chi)$ and $e(\chi)$, it fails to account for secular
changes in $w(\chi)$, $t(\chi)$, and $\phi(\chi)$; this conclusion
confirms and extends those of our previous work 
\cite{our_paper, other_paper}.

In addition, the radiative approximation is subject to the same
ambiguities regarding the choice of initial conditions as the
secular approximation. Writing the radiative evolution as the sum of
its secular and oscillatory parts, $I^A_{\mathrm{r}}(\chi) = 
I^A_{\mathrm{r\ sec}}+I^A_{\mathrm{r\ osc}}$, we shall consider three 
possible candidates for $I^A_{\mathrm{r}}(0)$. The first is 
$I^A_{\mathrm{r}}(0)=I^A(0)$, the exact initial data that is selected
for the true evolution of the orbital elements under the action of the
full perturbing force. The second is 
$I^A_{\mathrm{r\ sec}}=\av{I^A}_\chi(0)$, the $\chi$-averaged initial
data, which identifies the initial secular part of the radiative
evolution with the initial $\chi$-averaged part of the true
evolution. The third choice is 
$I^A_{\mathrm{r\ sec}}(0)=\av{I^A}_t(0)$, the $t$-averaged initial
data, which identifies the initial secular part of the radiative
evolution with the initial $t$-averaged part of the true
evolution. (We note that for both the second and third choices of
initial conditions, the initial value $I^A_{\mathrm{r}}(0)$ is not
fixed by $I^A_{\mathrm{r\ sec}}(0)$ alone, since we also require the
initial value of $I^A_{\mathrm{r\ osc}}$. Although we do not have 
\textit{a priori} access to this oscillatory part, we can assign it an
approximate initial value based on the results of the radiative
evolution with exact initial conditions. This introduces a negligible
error, since the oscillations in the radiative evolution are extremely
small in practice.) These three choices of initial data are distinct,
and they lead to different evolutions. We shall see that the accuracy
of the evolution (relative to the true evolution) depends strongly on
the choice of initial data. 

\subsubsection{Orbital evolution}

A typical inspiral of interest for LISA will form in a highly
eccentric state. Over the course of the inspiral the system will emit
gravitational radiation carrying away energy and angular momentum,
shrinking and circularizing the orbit over time. Thus, the inspiral
will evolve from a highly eccentric orbit to a quasi-circular one,
and it will end in a rapid plunge. We shall now determine the validity 
of the radiative approximation for this class of orbits. Since our
perturbing force is valid only in the post-Newtonian regime, we always  
ensure that $v^2\lesssim 0.1$.  

The general limitations of the radiative approximation are
demonstrated in Fig.~\ref{orbits}, which displays the spatial
trajectories of a highly eccentric orbit and a quasi-circular orbit,
along with corresponding radiative approximations. In each case the
true and approximate orbits are terminated at identical final times,
at which point the radiative approximation lags behind the true
orbit. With a mass ratio of $\mu/M=0.01$, this dephasing of the two
orbits is noticeable after only fifteen radial cycles in the eccentric 
case, while several dozen revolutions are required in the quasi-circular
case. Since the dephasing is apparent before any non-geodesic
precession occurs, we interpret its cause to be conservative effects
in the time-dependence of the orbit. That is, the error in $t(\chi)$
dominates over the errors in $w(\chi)$ and $\phi(\chi)$, such that the
particle lies at the wrong spatial point at a given time, even before
$r(\chi)$ and $\phi(\chi)$ have deviated significantly from the true
orbit. 

For the plots in Fig.~\ref{orbits} we have chosen exact initial
conditions $I^A(0)$ for the approximate orbit. By choosing averaged
initial conditions we obtain better results in the eccentric case: as
shown in Fig.~\ref{av_orbit}, using time-averaged initial conditions
$\av{I^A}_t(0)$ eliminates the dephasing on the timescale of the
plot. Using $\chi$-averaged initial conditions $\av{I^A}_\chi(0)$ 
results in a smaller improvement, as we will discuss below. However,
in the quasi-circular case all initial conditions fare equally well.  

The evolution of the orbital elements over a complete inspiral,
beginning with the initial conditions of the eccentric orbit in
Fig.~\ref{orbits} and continuing to quasi-circularity, is displayed in
Fig.~\ref{plots_vs_t}. Insets in the plots display the same range of 
time covered by Fig.~\ref{orbits}. The orbit stops before the final
plunge of the small body into the large black hole. There are two
reasons for this truncation. First, our method of osculating orbits
cannot cover the final plunge, because of the underlying restriction
that the orbit must be bounded between a minimum radius $pM/(1+e)$ and
a maximum radius $pM/(1-e)$; this is reflected mathematically by the
condition $p > 6+2e$, which is violated during plunge. Second, we
should in any case leave this portion of the orbit alone, because the
velocities and fields therein are highly relativistic; in this regime
the post-Newtonian expansion of the perturbing force becomes
inaccurate. In Fig.~\ref{plots_vs_t} we display results of the
numerical evolution for the principal element $p$ and positional
element $w$ only; the evolution of $e$ is qualitatively similar to
that of $p$. It is worth noting, however, that the eccentricity never
quite reaches $e \approx 0$; instead, quasi-circularity is manifested
by the condition $\chi - w \approx 0$, which equally well ensures that
$r'\approx 0$. This observation agrees with the results of
Ref.~\cite{Lincoln}.   

The results for all three choices of initial conditions are
plotted in Fig.~\ref{plots_vs_t}. As we see from these plots, the
radiative approximation qualitatively matches the true secular
evolution for the principal element $p$, but neglects all secular
changes in the positional element $w$. This is the expected
result. However, we also see that the radiative approximation deviates
from the true evolution even for the principal element. The extent of
this deviation depends on the choice of initial conditions, with the
time-averaged initial conditions faring the best and exact initial
conditions the worst.

An essential aspect of our results is that the errors in the principal
elements produced by the radiative approximation are mostly due to
errors in $t(\chi)$. As we see in Fig.~\ref{plots_vs_chi}, the errors
almost completely vanish when the principal elements are plotted as
functions of $\chi$; significant errors arise only in the conversion
between $\chi$ and $t$.

\begin{figure}[t]
\begin{center}
\includegraphics{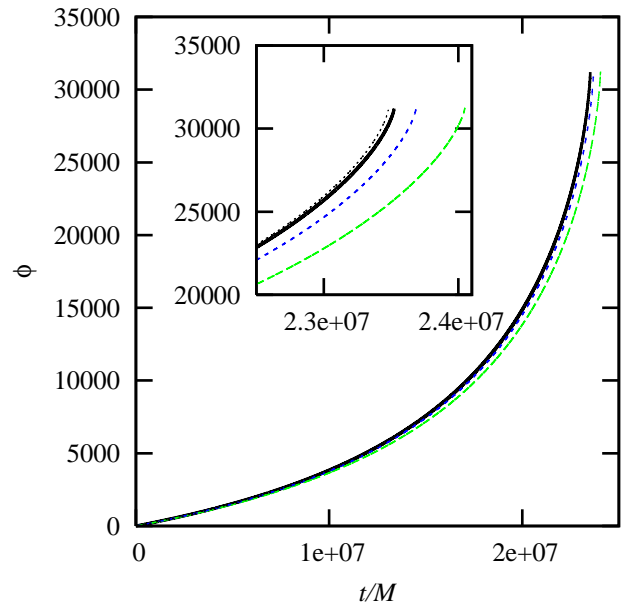}
\end{center}
\caption{The orbital phase $\phi$, with curves as described in
  Fig.~\ref{plots_vs_t}. The scale of the plot suggests that the
  radiative evolution with time-averaged initial data (the uppermost
  curve in dotted black) gives an accurate approximation of the
  true evolution. The vertical scale, however, is large, and this is a
  false impression. At the late time $t/M = 2.345\times 10^7$, the
  error in phase is $\Delta \phi = 4520$ rad for the exact initial
  data, $\Delta \phi = 1830$ rad for the $\chi$-averaged initial
  data, and $\Delta \phi = 655$ rad for the $t$-averaged initial
  data. This last choice fares best, but its accuracy is poor over a
  complete inspiral.}  
\label{phi_vs_t}
\end{figure}

\subsubsection{Errors in orbital phase}

The errors in which we are most interested are errors in orbital
phase, since they will lead directly to errors in the phase of the
emitted gravitational radiation. Figure~\ref{phi_vs_t} displays the
phase $\phi$ versus time, again using all three choices of initial
conditions for the radiative approximation. Once again we see that the
time-averaged conditions produce the smallest error, for the same
reasons described in the previous section. 

\begin{figure}[h!]
\begin{center}
\includegraphics[angle=-90]{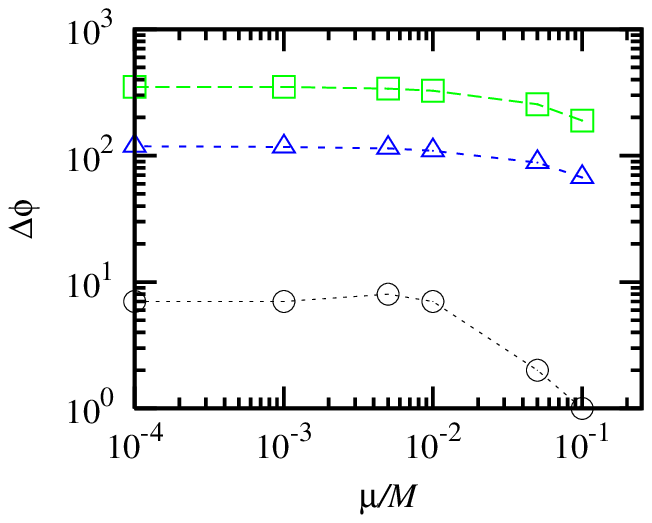}
\includegraphics[angle=-90]{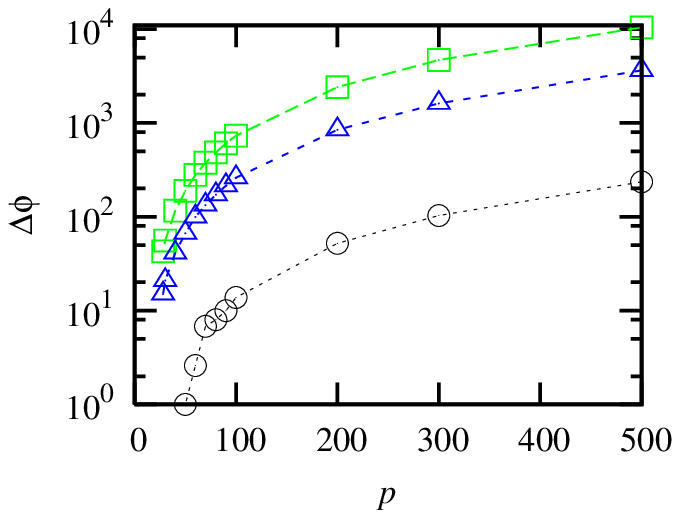}
\includegraphics[angle=-90]{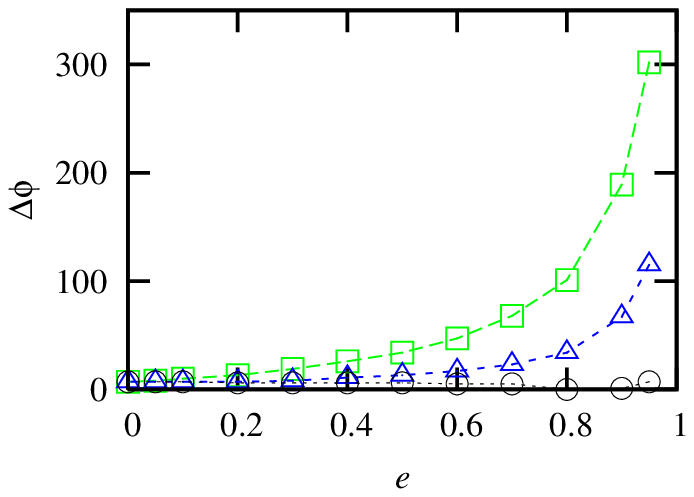}
\end{center}
\caption{The difference in orbital phase $\phi$ between the true orbit
  and approximate orbits after a radiation-reaction time (defined by
  $p\to 0.9p_0$ on the true orbit). Open squares indicate results for
  identical initial conditions, open triangles for matching
  $\chi$-averaged initial conditions, and open circles for matching
  $t$-averaged initial conditions. Top: dephasing as a function of the
  mass ratio $\mu/M$, with fixed true initial values $p_0=50$ and
  $e_0=0.9$. The dephasing becomes $\mu$-independent for sufficiently
  small $\mu$, when second-order effects become negligible. Middle:
  dephasing as a function of initial value $p_0$, with fixed $e_0=0.9$
  and $\mu/M=0.1$. Bottom: dephasing as a function of initial
  eccentricity $e_0$, with fixed $p_0=50$ and $\mu/M=0.1$. The error
  in the case of time-averaged initial conditions is approximately
  independent of $e$.} 
\label{dephasing}
\end{figure}

Figure~\ref{dephasing} shows the dependence of the dephasing
$\Delta\phi=\phi-\phi_{\mathrm{rad}}$ on the parameters of the
problem. We plot the dephasing for a ``radiation-reaction'' time
defined by $p\to 0.9p_0$, rather than a complete inspiral, since
gravitational-wave data analysis may  
require only a portion of a complete inspiral. We see that the
dephasing is independent of $\mu$ for sufficiently small values of
$\mu$. This is an expected result, since the radiation-reaction time
at leading order in $\mu$ varies as $1/\mu$, while the rate of
dephasing varies as $\mu$, leading to a net cancellation in the
total dephasing. However, terms in the perturbing force that are
quadratic in $\mu$ alter this result when $\mu/M$ is sufficiently
large. Somewhat surprisingly, these quadratic terms actually serve to
decrease the dephasing, lowering the impact of conservative terms in
the force.

As expected, the dephasing decreases at lower values of $e$, although 
the eccentricity seems to have negligible impact in the case of
time-averaged initial conditions. Also as expected, the dephasing
varies as $p^{3/2}$, regardless of initial conditions. This scaling
follows from the form of the post-Newtonian force: the leading-order
conservative term enters at 1PN order, which scales as a $p^{-1}$
correction to Newtonian gravitation, while the leading-order
dissipative term enters at 2.5PN order, which scales as a
$p^{-5/2}$ correction. The dephasing is governed by the relative
strength of the conservative terms, leading to a scaling of
$p^{-1}/p^{-5/2}=p^{3/2}$.    

In all cases the time-averaged initial conditions yield the best
results. Indeed, the efficacy of these initial conditions is almost
surprising. One way of understanding their impact is to examine the
insets in Fig.~\ref{plots_vs_t}. Peaks in the true curve correspond to
the short periods of time near periapsis, while relatively flat
regions correspond to the long periods of time 
around apoapsis. Thus, choosing exact initial conditions matches the
true and approximate orbits for the minimal amount of time, as well as
in the region of strongest fields, leading to the largest possible
deviation. Choosing time-averaged initial conditions matches the
orbits near apoapsis, for the longest time and with the weakest
fields, leading to the least possible deviation. The $\chi$-averaged
initial conditions are then in some sense the average of all the
incorrect choices. An implication of this is that in some
circumstances the $\chi$-averaged initial conditions could turn out
to be even worse than the exact initial conditions. For example,
choosing exact initial conditions at apoapsis would closely
approximate the time-averaged initial conditions, which would then
fare much better than the $\chi$-averaged initial conditions.

We can explain the long-term impact of the initial conditions by
considering the time-dependence of an orbit. The secular time function 
$\av{t}(\chi)$ can be written in terms of the orbital period 
$P(\chi)=\int_{\chi-\pi}^{\chi+\pi}t'(\tilde\chi)d\tilde\chi$
as $\av{t}(\chi)=\int_0^\chi P(\tilde\chi)d\tilde\chi$. As we see
from the insets in Fig.~\ref{plots_vs_t}, the changes in initial
conditions bring the initial orbital period of the radiative
approximation closer to that of the true orbit; and as we would
intuitively expect, the time-averaged initial conditions best
reproduce the initial temporal period. This correction, $\delta P$,
to the initial period then induces a long-term correction to
$\av{t}(\chi)$ of the form $\delta\av{t}\sim\chi\cdot\delta P$. (Such
an effect can be calculated explicitly for the electromagnetic
self-force considered in Ref.~\cite{other_paper}: refer to Eqs.~(4.8)
and (4.21) therein.) In essence, the time-averaged initial conditions
carry information about the initial conservative correction to the
true orbital period, and they thus implicitly insert a conservative
correction into the radiative approximation. This serves to remind us 
that we would have difficulty choosing suitable initial conditions
for the radiative approximation if we did not have prior access to
the true evolution. 

Regardless of the choice of initial conditions, one should note that
the errors accumulated over a complete inspiral are much larger than
those shown in Fig.~\ref{dephasing}. (Refer to the caption of 
Fig.~\ref{phi_vs_t} for actual values.) Also, the plots of
$\Delta\phi$ versus $p$ and $e$ are for $\epsilon=0.1$, leading to a
smaller dephasing than would occur if $\epsilon$ were in the region of
linear dominance. Thus, even if ideal initial conditions could be
found without reference to the exact solution, the radiative
approximation would generically fail over a complete inspiral.

\subsubsection{Gauge dependence}

%
As is well known, the gravitational self-force is a gauge-dependent
quantity: it is not invariant under a change of coordinates
$x^\mu \to x^\mu + \xi^\mu$, where $\xi^\mu$ is a ``small'' vector
field. The equations of motion that we have used in this paper were
calculated within the harmonic gauge of post-Newtonian theory, and the
magnitudes of the conservative effects that we have displayed refer to
this particular gauge choice; different gauges would necessarily lead
to different results. Indeed, Mino has argued in favor of constructing
a physically meaningful ``radiation-reaction gauge'' in which the
conservative effects of the self-force are set to zero over a finite
radiation-reaction time, making the radiative approximation exact over 
that interval \cite{Mino2005, Mino2006}. Mino has also argued that
this gauge choice induces a change in initial conditions that
partially absorbs conservative effects \cite{Mino2005}, and this
statement agrees with our result that long-term conservative effects
can be mimicked by a small change in initial conditions. We would like
to point out, however, that a rigorous construction and implementation
of such a gauge choice have yet to be performed, and that the impact
of making this choice on quantities other than the self-force has yet
to be determined. 

It is known, for example, that in the harmonic gauge of post-Newtonian 
theory, the equations of motion contain both radiative and
conservative terms, and that the gravitational potentials are
well-behaved everywhere, except at the position of each (pointlike)
body where they diverge with an expected power of $m/r$. What is the
behavior of the gravitational potentials in Mino's radiation-reaction
gauge? The answer is not known, and it would be interesting to
investigate the issue in post-Newtonian theory. For example, one could
determine the effect on the potentials of making a coordinate
transformation that would turn off some of the conservative terms in
the equations of motion (those that depend on $\epsilon$ in
Sec.~IIIA); would this spoil the behavior of the potentials near the
bodies, or perhaps elsewhere in the spacetime? Such an analysis would
be revealing, and it would give indication as to whether Mino's scheme
is likely to be successfully implemented.   

We believe that the Lorenz gauge of the gravitational self-force
problem, which is in close mathematical analogy with the harmonic
gauge of post-Newtonian theory, is also in close physical analogy: it
produces conservative terms in the self-force, and it produces
gravitational potentials that are well behaved everywhere (except at
the position of the orbiting body). Given the successes of
post-Newtonian theory in its harmonic-gauge formulation, we feel 
confident that the Lorenz gauge is ultimately a better choice of gauge 
for the gravitational self-force problem, in spite of the presence of
conservative terms in the equations of motion. We shall therefore
defer our judgment on the advantages of Mino's radiation-reaction
gauge, and reiterate the importance of the conservative terms in the
harmonic-gauge (or Lorenz-gauge) self-force. Our conclusions, to be
sure, apply within the confines of the post-Newtonian harmonic
gauge. But we contend that our conclusions are in fact generic:
Outside of a finely-tuned gauge choice, one should expect the
conservative part of the self-force to produce large secular effects.  
%


\section{Conclusion}

The first part of this paper was devoted to the development of a
method of osculating orbits to integrate the equations of motion that
govern bound, accelerated orbits in Schwarzschild spacetime. The
method involves the phase-space variables $\{p,e,w,t,\phi\}$, which
are expressed as functions of an orbital parameter $\chi$; each
variable satisfies a first-order differential equation, and knowledge
of these variables is sufficient to determine the worldline in
spacetime. Although the method is limited to situations in which the
force acts within the orbital plane, this limitation can be 
overcome; in addition, the force is not assumed to be small. We show
in Appendix \ref{Newtonian} that for large values of $p$, our
equations reduce to the standard perturbation equations of
Newtonian celestial mechanics. The method has many potential
applications, including the important one of permitting an
implementation of the gravitational self-force. Most immediately,
it provides an attractive conceptual and mathematical foundation for a
perturbative approach to weakly accelerated orbits. And furthermore,
the method is easy to implement in practice in a numerical code.    

In the second part of the paper we applied the method of osculating
orbits to the inspiral of a small body into a Schwarzschild black hole
of much larger mass. The perturbing force was calculated on the basis
of the hybrid Schwarzschild/post-Newtonian equations of motion of
Kidder, Will, and Wiseman \cite{Kidder}, and its effect on the
orbiting body was obtained by numerical integration of our evolution
equations for the dynamical variables $\{p,e,w,t,\phi\}$. This
approach is well suited to a study of the limitations and ambiguities
of adiabatic and radiative approximations, which was carried out
next. Specifically, we have illustrated the importance of conservative  
effects in the time dependence of the orbit, and we have established
the advantage of choosing time-averaged initial conditions for the
approximated orbital elements. This problem differs in many respects
from the fully relativistic self-force problem, but it nevertheless
captures many of its essential features. Our conclusions, therefore,
might be expected to hold in the fully relativistic case for most choices of gauge. 


\begin{acknowledgments}
We wish to thank the referee for several helpful suggestions. This
work was supported by the Natural Sciences and Engineering Research
Council of Canada.  
\end{acknowledgments}


\appendix
\section{Newtonian limit}
\label{Newtonian}

Since our work extends the standard methods of Newtonian celestial
mechanics, it is a worthwhile endeavor to show that our equations
reduce to those for perturbed Keplerian orbits in Newtonian
mechanics. In this Appendix we derive the Newtonian limit of our
expressions by expanding in powers $p^{-1}$; since $p^{-1} \propto
r^{-1} \sim v^2$, this is equivalent to a post-Newtonian expansion. We
shall first describe the general relationship between the Newtonian
and relativistic perturbing forces. Next we shall show that our
geodesic parametrization reduces to Keplerian ellipses and that our
evolution equations for the orbital elements $p$, $e$, and $w$ reduce
to Gauss' perturbation equations of celestial mechanics.  

Substituting the Christoffel symbols of the Schwarzschild metric into
the equations of motion~\eqref{affine eq mot} yields the following
equations for the force: 
\begin{eqnarray}
f^r & = & \ddot r +F\frac{M}{r^2}\dot t^2 
- F^{-1}\frac{M}{r^2}\dot r^2+F\dot\phi^2, \\
f^{\phi}&=&\ddot\phi+2\frac{\dot r\dot\phi}{r}, \\ 
f^t & = & \ddot t + F^{-1}\frac{2M}{r^2}\dot{r}^2, 
\end{eqnarray}
where $F = 1-2M/r$. The time-component of the force can be written in
a more useful form using the orthogonality
relation~\eqref{orthogonality}.  

These expressions for the relativistic force differ nontrivially from
those in the Newtonian case. We define $\tilde F$, the Newtonian
perturbing force per unit mass, via Newton's second law: 
\begin{equation}
\ddot{\bm{x}} = \bm{g}+\tilde{\bm{F}},
\end{equation}
where $\bm{x}$ is a 3-vector representing the spatial coordinates of
the particle and $\bm{g}=-\frac{M}{r^2}\hat{\bm{r}}$ is the Newtonian
gravitational acceleration. For convenience we have defined the
Newtonian acceleration as the second derivative of $\bm{x}$ with
respect to proper time rather than coordinate time. We also define the
radial and tangential components of the perturbing force via
\begin{equation} 
\tilde{\bm{F}}\equiv \tilde F^r \hat{\bm{r}} + 
\tilde F^{\phi} \hat{\bm{\phi}}, 
\end{equation} 
where $\hat{\bm{r}}$ and $\hat{\bm{\phi}}$ form an orthonormal basis
in the orbital plane. Given these definitions, writing $\ddot{\bm{x}}$
in polar coordinates $(r,\phi)$ leads to  
\begin{eqnarray}
\tilde F^r & = & \ddot r - r\dot\phi^2+\frac{M}{r^2} \\
\tilde F^{\phi} & = & r\ddot\phi +2\dot r\dot\phi.
\end{eqnarray}

Comparing the Newtonian and relativistic expressions for the
perturbing force, we see they are related by the equations 
\begin{eqnarray}
f^r & = & \tilde F^r + r\left(1-F\right)\dot\phi^2 \nonumber\\ 
& & +\frac{M}{r^2}\left(F\dot{t}^2+F^{-1}\dot{r}^2-1\right), 
\label{fr relation}\\
f^{\phi} & = & \frac{\tilde F^{\phi}}{r}\label{fphi relation}. 
\end{eqnarray}
Thus, $f^r$ differs from $\tilde F^r$ by relativistic corrections,
while $f^{\phi}$ differs from $\tilde F^{\phi}$ only by a factor of
the orbital radius. 

We next consider our parametrization of geodesics. From
Eqs.~\eqref{phiprime}, \eqref{tprime}, and \eqref{chidot} one
trivially finds the leading-order terms in $\phi'$, $t'$, and
$\dot\chi$ to be  
\begin{eqnarray}
\phi' & = & 1, \label{phiprime expansion}\\
t' & = & \frac{p^{3/2}M}{[1+e\cos(\chi-w)]^2}, 
\label{tprime expansion}\\
\dot\chi & = & \frac{[1+e\cos(\chi-w)]^2}{p^{3/2}M}.
\label{chidot expansion}
\end{eqnarray}
Thus, in the Newtonian limit we have $\phi=\chi$ and $t=\tau$ and the
resulting parametrization 
\begin{eqnarray}
r & = & \frac{pM}{1+e\cos(\phi-w)},\\
\diff{\phi}{t} & = & \frac{[1+e\cos(\phi-w)]^2}{p^{3/2}M}.
\end{eqnarray}

In terms of the orbital elements, we see that $w=\Phi$ in the
Newtonian limit. This corresponds to the loss of one degree
of freedom, as we would expect from the fact that $t$ in Newtonian
physics is a universal parameter rather than a coordinate. We can also 
easily find that the energy and angular momentum per unit mass reduce
to $E=1-\frac{1-e^2}{2p}$ and $L=\sqrt{p}M$, respectively. The first
term in $E$ is the rest energy of the particle, while the second term
is the Newtonian energy $\frac{1}{2}v^2-\frac{M}{r}$. 

With the exception of the inclusion of the rest mass, the above
results are standard Keplerian relationships. Thus, our equations for
the orbital elements should reduce to those for perturbed Keplerian
orbits. Substituting Eqs.~\eqref{phiprime expansion}--\eqref{chidot
 expansion} into Eqs.~\eqref{orthogonality}, \eqref{fr relation}, and
\eqref{fphi relation}, we find the leading-order expressions for the
perturbing force: 
\begin{eqnarray}
f^r & = & \tilde F^r \\
f^{\phi}&=&\frac{\tilde F^{\phi}}{r} \\
f^t & = & \frac{e\sin(\phi-w)}{\sqrt{p}}\tilde{F}^r 
+ \frac{1+e\cos(\phi-w)}{\sqrt{p}}\tilde{F}^{\phi}. \qquad
\end{eqnarray}
These results allow us to expand Eqs.~\eqref{pprime}, \eqref{eprime},
and \eqref{wprime} to find the leading-order expressions for the
orbital elements: 
\begin{eqnarray}
\diff{p}{t} & = & \frac{2p^{3/2}}{1+e\cos(\phi-w)}\tilde F^{\phi} \\ 
\diff{e}{t} & = & \sqrt{p}\ \frac{e+2\cos(\phi-w)+e\cos^2(\phi-w)}
{1+e\cos(\phi-w)}\tilde F^{\phi} \nonumber\\
&& +\sqrt{p}\ \sin(\phi-w)\tilde F^r\\
\diff{w}{t} & = & \frac{\sqrt{p}M^{3/2}}{e}\ 
\frac{\sin(\phi-w)[2+e\cos(\phi-w)]}{1+e\cos(\phi-w)}\tilde{F}^{\phi} 
\nonumber\\
&& -\frac{\sqrt{p}M^{3/2}}{e}\ \cos(\phi-w)\tilde{F}^r.
\end{eqnarray}
These are Gauss' well known perturbation equations.


\section{Evolution equations from Killing vectors}
\label{elements from E and L}

It is possible to derive Eqs.~\eqref{pprime}--\eqref{wprime} for the 
derivatives of the osculating elements from Eq.~\eqref{diffI 1b} and
the Killing vectors of the Schwarzschild spacetime, without reference
to Eq.~\eqref{diffI 2b}. Although this derivation is equivalent to
that given in Sec.~\ref{evolution}, its physical significance is more
intuitive. We begin by defining energy and angular momentum (per unit
mass) as $E = -\xi_{(t)}^{\alpha}\dot{z}_{\alpha}$ and
$L = \xi_{(\phi)}^{\alpha}\dot{z}_{\alpha}$, where
$\xi_{(t)}=\pdiff{}{t}$ and $\xi_{(\phi)}=\pdiff{}{\phi}$ are Killing
vectors corresponding to the spacetime's invariance under time
translations and spatial rotations. From these definitions we find  
\begin{eqnarray}
-\dot{E} & = &
\dot{z}^{\beta}(\xi_{(t)}^{\alpha}\dot{z}_{\alpha})_{;\beta} 
\nonumber\\
& = & \xi^{\alpha}_{(t);\beta}\dot{z}_{\alpha}\dot{z}^{\beta}
+ \xi^{\alpha}_{(t)}\dot{z}^{\beta}\dot{z}_{\alpha;\beta} \nonumber\\
& = & \xi^{\alpha}_{(t)}f_{\alpha}.
\end{eqnarray}
The first term on the second line vanishes due to the antisymmetry of
$\xi_{\alpha;\beta}$ for any Killing vector $\xi$, and the final line
then follows from the equation of motion
$\dot{z}^{\alpha}\dot{z}^{\beta}{}_{;\alpha}=f^{\beta}$. An analogous
result holds for $\dot L$. From the definitions of $\xi_{(t)}$ and
$\xi_{(\phi)}$ we then find 
\begin{eqnarray}
\dot E & = & Ff^t, \\
\dot L & = & r^2f^{\phi}.
\end{eqnarray}

These results can be used to find $\dot e$ and $\dot p$ using 
Eqs.~\eqref{E} and \eqref{L}, which define $E(p,e)$ and
$L(p,e)$. Using these relationships, we write $\dot E =
\pdiff{E}{p}\dot p + \pdiff{E}{e}\dot e$ and $\dot L =
\pdiff{L}{p}\dot p + \pdiff{L}{e}\dot e$, which can be rearranged to 
find 
\begin{eqnarray} 
\dot p & = & \frac{\pdiff{E}{e}\dot L 
- \pdiff{L}{e}\dot E}{\pdiff{L}{p}\pdiff{E}{e}
- \pdiff{L}{e}\pdiff{E}{p}}, \\
\dot e & = & \frac{\pdiff{L}{p}\dot E 
- \pdiff{E}{p}\dot L}{\pdiff{L}{p}\pdiff{E}{e}
- \pdiff{L}{e}\pdiff{E}{p}}.
\end{eqnarray}
The equation for $\dot w$ can then be found from Eq.~\eqref{diffI 1b},
which leads to Eq.~(\ref{wprime_raw}), or 
\begin{equation}
\dot w = \frac{1}{r'}\left(\pdiff{r}{e}\dot e 
+ \pdiff{r}{p}\dot p\right).
\end{equation}

The explicit results of these calculations are
\begin{widetext}
\begin{eqnarray}
\dot p & = & -\frac{2p^{1/2}(p-2-2e\cos v)(p-2-2e)^{1/2}(p-2+2e)^{1/2}
(p-3-e^2)^{1/2}}{(p-6+2e)(p-6-2e)}f^t\nonumber\\
\nonumber\\
&&+\frac{2p^2M(p-4)^4(p-3-e^2)^{1/2}}{(p-6+2e)(p-6-2e)
(1+e\cos v)^2}f^{\phi}, \\
\nonumber\\
\dot e & = & \frac{(p-6-2e^2)(p-2-2e\cos v)(p-2-2e)^{1/2}
(p-2+2e)^{1/2}(p-3-e^2)^{1/2}}{p^{1/2}e(p-6+2e)(p-6-2e)}f^t
\nonumber\\
\nonumber\\
&&-\frac{pM(1-e^2)(p^2-8p+12+4e^2)(p-3-e^2)^{1/2}}{e(p-6+2e)
(p-6-2e)(1+e\cos v)^2}f^{\phi},  \\
\nonumber\\
\dot w & = & -\frac{(2e+(p-6)\cos v)(p-2-2e\cos v)(p-2-2e)^{1/2}
(p-2+2e)^{1/2}(p-3-e^2)^{1/2}} {p^{1/2}e^2\sin v(p-6+2e)(p-6-2e)}f^t
\nonumber\\
&&+\frac{pM\lbrace2e(p^2-8p+32)+[(p^2-8p)(1+e^2)+4e^2(6-e^4)]
\cos v\rbrace(p-3-e^2)^{1/2}} 
{e^2\sin v(p-6+2e)(p-6-2e)(1+e\cos v)^2}f^{\phi}.
\end{eqnarray}
\end{widetext}
When accompanied by the auxiliary equation \eqref{chidot} for
$\diff{\chi}{\tau}$, these equations form a closed, autonomous system
for the orbital elements. 

The results in this section are equivalent to those in
Sec.~\ref{evolution}, which can be easily shown by using
Eq.~\eqref{orthogonality} to replace $f^t$ with $f^r$. But they are
numerically ill-behaved. Specifically, $\dot e$ appears to diverge in
the limit $e\to 0$, and $\dot w$ appears to diverge when $\sin v = 0$
(i.e., at every turning point in the orbit). Although these divergences
are canceled analytically by the numerators in each case, they are
serious obstacles in a numerical integration. Thus, the equations
given in Sec.~\ref{evolution} are more practical, though slightly
lengthier.  

\bibliography{osculating}

\begin{thebibliography}{21}
\expandafter\ifx\csname natexlab\endcsname\relax\def\natexlab#1{#1}\fi
\expandafter\ifx\csname bibnamefont\endcsname\relax
  \def\bibnamefont#1{#1}\fi
\expandafter\ifx\csname bibfnamefont\endcsname\relax
  \def\bibfnamefont#1{#1}\fi
\expandafter\ifx\csname citenamefont\endcsname\relax
  \def\citenamefont#1{#1}\fi
\expandafter\ifx\csname url\endcsname\relax
  \def\url#1{\texttt{#1}}\fi
\expandafter\ifx\csname urlprefix\endcsname\relax\def\urlprefix{URL }\fi
\providecommand{\bibinfo}[2]{#2}
\providecommand{\eprint}[2][]{\url{#2}}

\bibitem[{\citenamefont{Damour}(1987)}]{300_years}
\bibinfo{author}{\bibfnamefont{T.}~\bibnamefont{Damour}},
  \emph{\bibinfo{title}{300 Years of Gravitation}}
  (\bibinfo{publisher}{Cambridge University Press},
  \bibinfo{address}{Cambridge, England}, \bibinfo{year}{1987}), pp.
  \bibinfo{pages}{128--198}.

\bibitem[{\citenamefont{Futamase and Itoh}(2007)}]{PN_review2}
\bibinfo{author}{\bibfnamefont{T.}~\bibnamefont{Futamase}} \bibnamefont{and}
  \bibinfo{author}{\bibfnamefont{Y.}~\bibnamefont{Itoh}},
  \bibinfo{journal}{Living Rev. Rel.} \textbf{\bibinfo{volume}{10}}
  (\bibinfo{year}{2007}), \bibinfo{note}{[Online article]: cited on \today},
  \eprint{http://www.livingreviews.org/lrr-2007-2}.

\bibitem[{\citenamefont{Blanchet}(2006)}]{Blanchet}
\bibinfo{author}{\bibfnamefont{L.}~\bibnamefont{Blanchet}},
  \bibinfo{journal}{Living Rev. Relativity} \textbf{\bibinfo{volume}{9}}
  (\bibinfo{year}{2006}), \bibinfo{note}{[Online article]:
  http://www.livingreviews.org/lrr-2006-4}.

\bibitem[{LIS()}]{LISA}
\bibinfo{note}{The LISA website is located at http://lisa.jpl.nasa.gov}.

\bibitem[{\citenamefont{Mino et~al.}(1997)\citenamefont{Mino, Sasaki, and
  Tanaka}}]{Mino_etal}
\bibinfo{author}{\bibfnamefont{Y.}~\bibnamefont{Mino}},
  \bibinfo{author}{\bibfnamefont{M.}~\bibnamefont{Sasaki}}, \bibnamefont{and}
  \bibinfo{author}{\bibfnamefont{T.}~\bibnamefont{Tanaka}},
  \bibinfo{journal}{Phys. Rev. D} \textbf{\bibinfo{volume}{55}},
  \bibinfo{pages}{3457} (\bibinfo{year}{1997}),
  \bibinfo{note}{arXiv:gr-qc/9606018}.

\bibitem[{\citenamefont{Quinn and Wald}(1997)}]{QuinnWald}
\bibinfo{author}{\bibfnamefont{T.~C.} \bibnamefont{Quinn}} \bibnamefont{and}
  \bibinfo{author}{\bibfnamefont{R.~M.} \bibnamefont{Wald}},
  \bibinfo{journal}{Phys. Rev. D} \textbf{\bibinfo{volume}{56}},
  \bibinfo{pages}{3381} (\bibinfo{year}{1997}),
  \bibinfo{note}{arXiv:gr-qc/9610053}.

\bibitem[{\citenamefont{Poisson}(2004)}]{self_force_review}
\bibinfo{author}{\bibfnamefont{E.}~\bibnamefont{Poisson}},
  \bibinfo{journal}{Living Rev. Rel.} \textbf{\bibinfo{volume}{7}}
  (\bibinfo{year}{2004}), \bibinfo{note}{[Online article]: cited on \today},
  \eprint{http://www.livingreviews.org/lrr-2004-6}.

\bibitem[{\citenamefont{Drasco}(2006)}]{Drasco_review}
\bibinfo{author}{\bibfnamefont{S.}~\bibnamefont{Drasco}},
  \bibinfo{journal}{Class. Quant. Grav.} \textbf{\bibinfo{volume}{23}},
  \bibinfo{pages}{S769} (\bibinfo{year}{2006}), \eprint{gr-qc/0604115}.

\bibitem[{\citenamefont{Taff}(1985)}]{Taff}
\bibinfo{author}{\bibfnamefont{L.}~\bibnamefont{Taff}},
  \emph{\bibinfo{title}{Celestial Mechanics: A Computational Guide for the
  Practitioner}} (\bibinfo{publisher}{John Wiley \& Sons},
  \bibinfo{address}{New York}, \bibinfo{year}{1985}).

\bibitem[{\citenamefont{Beutler}(2005)}]{Beutler}
\bibinfo{author}{\bibfnamefont{G.}~\bibnamefont{Beutler}},
  \emph{\bibinfo{title}{Methods of Celestial Mechanics}} (\bibinfo{publisher}{Springer},
  \bibinfo{address}{New York}, \bibinfo{year}{2005}).

\bibitem[{\citenamefont{Damour and Deruelle}(1985)}]{Damour}
\bibinfo{author}{\bibfnamefont{T.}~\bibnamefont{Damour}} \bibnamefont{and}
  \bibinfo{author}{\bibfnamefont{N.}~\bibnamefont{Deruelle}},
  \bibinfo{journal}{Ann. Inst. Henri Poincar\'e} \textbf{\bibinfo{volume}{43}},
  \bibinfo{pages}{107} (\bibinfo{year}{1985}).

\bibitem[{\citenamefont{Damour}(2004)}]{Damour2004}
\bibinfo{author}{\bibfnamefont{T.}~\bibnamefont{Damour}},
  \bibinfo{author}{\bibfnamefont{A.}~\bibnamefont{Gopakumar}}, \bibnamefont{and}
  \bibinfo{author}{\bibfnamefont{B.~R.}~\bibnamefont{Iyer}},
  \bibinfo{journal}{Phys. Rev. D} \textbf{\bibinfo{volume}{70}},
  \bibinfo{pages}{064028} (\bibinfo{year}{2004}),
  \bibinfo{note}{arXiv:gr-qc/0404128}.

\bibitem[{\citenamefont{Mino}(2005)}]{Mino2005}
\bibinfo{author}{\bibfnamefont{Y.}~\bibnamefont{Mino}}, \bibinfo{journal}{Prog. Theor. Phys.} \textbf{\bibinfo{volume}{113}}, \bibinfo{pages}{733-761}
  (\bibinfo{year}{2005}), \eprint{gr-qc/0506003}.

\bibitem[{\citenamefont{Kidder et~al.}(1993)\citenamefont{Kidder, Will, and
  Wiseman}}]{Kidder}
\bibinfo{author}{\bibfnamefont{L.~E.} \bibnamefont{Kidder}},
  \bibinfo{author}{\bibfnamefont{C.~M.} \bibnamefont{Will}}, \bibnamefont{and}
  \bibinfo{author}{\bibfnamefont{A.~G.} \bibnamefont{Wiseman}},
  \bibinfo{journal}{Phys. Rev. D} \textbf{\bibinfo{volume}{47}},
  \bibinfo{pages}{3281} (\bibinfo{year}{1993}).

\bibitem[{\citenamefont{Detweiler and Poisson}(2004)}]{Eric_Steve}
\bibinfo{author}{\bibfnamefont{S.}~\bibnamefont{Detweiler}} \bibnamefont{and}
  \bibinfo{author}{\bibfnamefont{E.}~\bibnamefont{Poisson}},
  \bibinfo{journal}{Phys. Rev.} \textbf{\bibinfo{volume}{D69}},
  \bibinfo{pages}{084019} (\bibinfo{year}{2004}), \eprint{gr-qc/0312010}.

\bibitem[{\citenamefont{Pound et~al.}(2005)\citenamefont{Pound, Poisson, and
  Nickel}}]{our_paper}
\bibinfo{author}{\bibfnamefont{A.}~\bibnamefont{Pound}},
  \bibinfo{author}{\bibfnamefont{E.}~\bibnamefont{Poisson}}, \bibnamefont{and}
  \bibinfo{author}{\bibfnamefont{B.~G.} \bibnamefont{Nickel}},
  \bibinfo{journal}{Phys. Rev. D} \textbf{\bibinfo{volume}{72}},
  \bibinfo{pages}{124001} (\bibinfo{year}{2005}), \eprint{gr-qc/0509122}.

\bibitem[{\citenamefont{Pound and Poisson}(2007)}]{other_paper}
\bibinfo{author}{\bibfnamefont{A.}~\bibnamefont{Pound}} \bibnamefont{and}
  \bibinfo{author}{\bibfnamefont{E.}~\bibnamefont{Poisson}}
  (\bibinfo{year}{2007}), \bibinfo{note}{in preparation}.

\bibitem[{\citenamefont{Mino}(2003)}]{Mino}
\bibinfo{author}{\bibfnamefont{Y.}~\bibnamefont{Mino}}, \bibinfo{journal}{Phys.
  Rev. D} \textbf{\bibinfo{volume}{67}}, \bibinfo{pages}{084027}
  (\bibinfo{year}{2003}), \eprint{gr-qc/0302075}.

\bibitem[{\citenamefont{Ganz et~al.}(2007)\citenamefont{Ganz, Hikida, Nakano,
  Sago, and Tanaka}}]{Nakano}
\bibinfo{author}{\bibfnamefont{K.}~\bibnamefont{Ganz}},
  \bibinfo{author}{\bibfnamefont{W.}~\bibnamefont{Hikida}},
  \bibinfo{author}{\bibfnamefont{H.}~\bibnamefont{Nakano}},
  \bibinfo{author}{\bibfnamefont{N.}~\bibnamefont{Sago}}, \bibnamefont{and}
  \bibinfo{author}{\bibfnamefont{T.}~\bibnamefont{Tanaka}}
  (\bibinfo{year}{2007}), \bibinfo{note}{to be published in Prog. Theor.
  Phys.}, \eprint{gr-qc/0702054}.

\bibitem[{\citenamefont{Sundararajan et~al.}(2007)\citenamefont{Sundararajan,
  Khanna, and Hughes}}]{Hughes}
\bibinfo{author}{\bibfnamefont{P.~A.} \bibnamefont{Sundararajan}},
  \bibinfo{author}{\bibfnamefont{G.}~\bibnamefont{Khanna}}, \bibnamefont{and}
  \bibinfo{author}{\bibfnamefont{S.~A.} \bibnamefont{Hughes}}
  (\bibinfo{year}{2007}), \bibinfo{note}{preprint}, \eprint{gr-qc/0703028}.

\bibitem[{\citenamefont{Pound}(2006)}]{my_thesis}
\bibinfo{author}{\bibfnamefont{A.}~\bibnamefont{Pound}}, Master's thesis,
  \bibinfo{school}{University of Guelph} (\bibinfo{year}{2006}).

\bibitem[{\citenamefont{Chandrasekhar}(1983)}]{Chandra}
\bibinfo{author}{\bibfnamefont{S.}~\bibnamefont{Chandrasekhar}},
  \emph{\bibinfo{title}{The Mathematical Theory of Black Holes}}
  (\bibinfo{publisher}{Oxford University Press}, \bibinfo{address}{New York},
  \bibinfo{year}{1983}).

\bibitem[{\citenamefont{Cutler et~al.}(1994)\citenamefont{Cutler, Kennefick,
  and Poisson}}]{parametrization}
\bibinfo{author}{\bibfnamefont{C.}~\bibnamefont{Cutler}},
  \bibinfo{author}{\bibfnamefont{D.}~\bibnamefont{Kennefick}},
  \bibnamefont{and} \bibinfo{author}{\bibfnamefont{E.}~\bibnamefont{Poisson}},
  \bibinfo{journal}{Phys. Rev. D} \textbf{\bibinfo{volume}{50}},
  \bibinfo{pages}{3816} (\bibinfo{year}{1994}).

\bibitem[{\citenamefont{Lincoln and Will}(1990)}]{Lincoln}
\bibinfo{author}{\bibfnamefont{C.~W.} \bibnamefont{Lincoln}} \bibnamefont{and}
  \bibinfo{author}{\bibfnamefont{C.~M.} \bibnamefont{Will}},
  \bibinfo{journal}{Phys. Rev. D} \textbf{\bibinfo{volume}{42}},
  \bibinfo{pages}{1123} (\bibinfo{year}{1990}).

\bibitem[{\citenamefont{Mino}(2006)}]{Mino2006}
\bibinfo{author}{\bibfnamefont{Y.}~\bibnamefont{Mino}}, \bibinfo{journal}{Prog. Theor. Phys.} \textbf{\bibinfo{volume}{115}}, \bibinfo{pages}{43-61}
  (\bibinfo{year}{2006}), \eprint{gr-qc/0601019}.

\end{thebibliography}

\end{document}